\documentclass[american,aps,pra,reprint,tightenlines,superscriptaddress,twocolumn,twoside,longbibliography]{revtex4-2}

\pdfoutput=1

\usepackage[utf8]{inputenc}
\usepackage{wrapfig}

\usepackage[unicode=true,pdfusetitle, bookmarks=true,bookmarksnumbered=false,bookmarksopen=false, breaklinks=false,pdfborder={0 0 0},backref=false,colorlinks=false] {hyperref}
\hypersetup{ colorlinks,linkcolor=myurlcolor,citecolor=myurlcolor,urlcolor=myurlcolor}
\usepackage{braket,colortbl,amsthm,amsmath,amssymb,dsfont}
\definecolor{myurlcolor}{rgb}{0,0,0.7}
\usepackage{graphics,graphicx,relsize}
\usepackage{color}
\usepackage{XCharter}
\usepackage[T1]{fontenc}

\usepackage{bm}

\usepackage{enumitem}  % Add this to your preamble

\setlist[itemize]{left=0pt, labelindent=0pt}

\usepackage{graphicx}
\usepackage{subcaption}
\usepackage{url}
\usepackage[font=small,labelfont=bf,justification=raggedright]{caption}

\usepackage{mathptmx}
\usepackage[left=1.6cm,right=1.6cm,top=2.45cm,bottom=2.45cm]{geometry}
\usepackage{bigints}
\usepackage{accents}

\usepackage{multirow}
\usepackage{multirow}
\usepackage[table,xcdraw]{xcolor}

\usepackage{float} % Add this in your preamble if not already present

\theoremstyle{plain}

\def\bea{\begin{eqnarray}}
\def\eea{\end{eqnarray}}
\def\ba{\begin{array}}
\def\ea{\end{array}}

\def\beq{\begin{equation}}
\def\eeq{\end{equation}}

\usepackage[normalem]{ulem}
 
\makeatletter

\begin{document}
\begingroup
\maketitle
\endgroup

\title{MATBG Josephson diode as an universal thermal machine}

\author{Hadi Mohammed Soufy}
\email{hm.soufy@niser.ac.in}
\affiliation{School of Physical Sciences, National Institute of Science Education and Research, HBNI, Jatni-752050, India}
\affiliation{Homi Bhabha National Institute, Training School Complex, Anushakti Nagar, Mumbai, 400094, India}

\author{Colin Benjamin}
\email{colin.nano@gmail.com}
\affiliation{School of Physical Sciences, National Institute of Science Education and Research, HBNI, Jatni-752050, India}
\affiliation{Homi Bhabha National Institute, Training School Complex, Anushakti Nagar, Mumbai, 400094, India}

%%%%%%%%%%%%%%%%%%%%%%%%%%%%%%%%%%%%%%%%%%%%%%%%%%%%%%%%%%%%%%%%%%%%%%%%%%%%%%%%%%%%%%%%%%%%%%%%%%
\begin{abstract}
Magic-angle twisted bilayer graphene Josephson junctions (MATBG-JJ) with a gate-tunable valley-polarized weak link exhibit an intrinsic Josephson diode effect originating from broken symmetries associated with valley polarization and band-structure anisotropy. Exploiting this nonreciprocal superconducting platform, we construct quantum Stirling (QSC), Otto (QOC), and Carnot (QCC) thermodynamic cycles, where the valley-polarization potential $\Delta_v$ acts as the principal control parameter, in contrast to conventional Josephson thermal machines driven by superconducting phase bias. We systematically compare the performance of MATBG-based Josephson diode thermal machines (MATBG-JDTM) with MATBG-based Josephson junction thermal machines (MATBG-JJTM) and AA-stacked bilayer graphene Josephson junction thermal machines (AABLG-JJTM). Owing to the flat-band-enhanced density of states and electrically tunable nonreciprocal transport in MATBG, both MATBG-JDTM and MATBG-JJTM exhibit significantly enhanced work output and efficiency over a broad operating regime compared to AABLG-JJTM. In particular, the gate-controlled MATBG-JDTM provides a flux-free alternative to conventional phase-driven architectures, mitigating limitations associated with magnetic-flux control and flux-noise effects. Our results establish MATBG Josephson diode platforms as a promising route toward electrically tunable quantum thermal machines and nonreciprocal superconducting caloritronics.
\end{abstract}
%%%%%%%%%%%%%%%%%%%%%%%%%%%%%%%%%%%%%%%%%%%%%%%%%%%%%%%%%%%%%%%%%%%%%%%%%%%%%%%%%%%%%%%%%%%%%%%%%%
\maketitle
\clearpage  % Ensures the introduction starts in the first column

\textbf{\textcolor{red}{Introduction:}} Josephson junction (JJ) devices have attracted sustained interest owing to their broad applications in superconducting electronics and quantum technologies, including ultra-sensitive detectors for electromagnetic radiation and single photons \cite{1josephson1962possible,2clarke1971electronics,3richards1977josephson}, nanoscale thermometry \cite{4giazotto2006opportunities}, micro-refrigeration and electronic cooling \cite{6leivo1996efficient,8nguyen2016cascade,10marchegiani2018chip}, and thermoelectric energy conversion \cite{11pekola2015towards}. More recently, JJs have emerged as superconducting diodes exhibiting the Josephson diode effect (JDE), where nonreciprocal supercurrents enable directional dissipationless transport \cite{12marchegiani2020nonlinear,14ando2020observation,15wu2022field}. Beyond charge transport, Josephson junctions provide a versatile platform for phase-coherent caloritronics, including thermal interferometers, heat valves, and thermal rectifiers \cite{16.5sothmann2017high,17guttman1997phase,19fornieri2017towards}, as well as quantum thermal machines such as heat engines and refrigerators operating through quantum thermodynamic cycles \cite{20campisi2015nonequilibrium,22marchegiani2016self,23pal2022josephson,25scharf2020topological}.

Magic-angle twisted bilayer graphene (MATBG), with its strongly correlated and symmetry-broken electronic phases, offers an ideal setting for realizing JJ-based devices with highly tunable functionalities \cite{26cao2018unconventional,27cao2018correlated,29bistritzer2011moire,31bistritzer2010transport}. Recent experiments on locally gated MATBG JJs have revealed unconventional Fraunhofer interference patterns together with the emergence of a JDE \cite{32de2021gate,33rodan2021highly,34diez2023symmetry,35rothstein2026gate}. These observations have stimulated extensive theoretical efforts to uncover the microscopic origin of the diode response, including symmetry breaking from unconventional superconducting pairing \cite{36alvarado2023intrinsic}, effects of large kinetic inductance and spatially inhomogeneous supercurrent distributions \cite{35rothstein2026gate}, and the interplay of trigonal Fermi-surface warping with valley polarization in weak-link regions \cite{37xie2023varphi,38hu2023josephson}.

\begin{figure}[!htbp]
    \centering
    \includegraphics[width=0.9\linewidth]{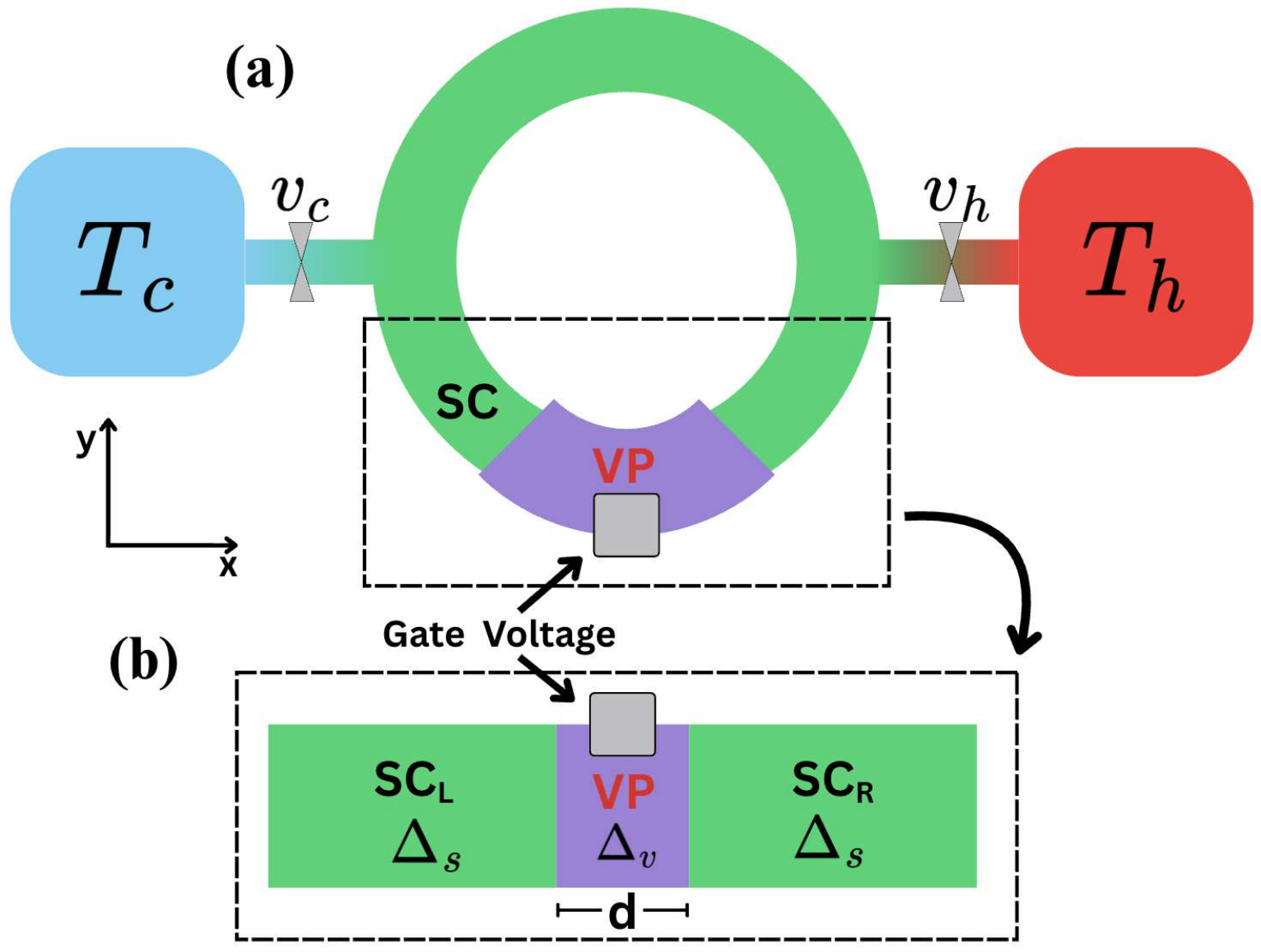}
    \caption{(a) Schematic of a MATBG-JDTM loop with a weak-link valley-polarized region of length $d$, coupled to thermal baths at temperatures $T_h$ and $T_c$ via controllable valves $v_h$ and $v_c$. (b) \(\text{SC}_\text{L}\) and \(\text{SC}_\text{R}\) are the left and right superconducting leads which are characterized by the superconducting order parameter $\Delta_s$, and \(\text{VP}\) is weakly-linked valley-polarized region characterized by potential $\Delta_v$ which is tunable by external gating. The phase difference across the junction is set to zero, since we are dealing with a Josephson diode (JD).}
    \label{fig:fig1}
\end{figure}

Previous studies have investigated MATBG-based thermal machines under perpendicular magnetic fields within quantum thermodynamic cycles \cite{39soufy2025enhanced,41soufy2026high}. In this Letter, we introduce MATBG-JDTM, where the valley-polarization potential serves as an electrically tunable thermodynamic control parameter in the absence of a superconducting phase bias. We systematically compare its performance with that of MATBG-JJTM and AABLG-JJTM, both driven by a finite superconducting phase difference without valley polarization. This comparison enables us to elucidate the advantages of purely electrical control over conventional phase-driven architectures relying on externally applied magnetic fields, while simultaneously revealing the crucial role of trigonal warping and the suppressed Fermi velocity intrinsic to MATBG in enhancing thermal-machine performance.

\textbf{\textcolor{red}{JDTM and JJTM :}} Fig.~\ref{fig:fig1}(a) illustrates a schematic of a JJ loop coupled to two thermal baths maintained at temperatures $T_c$ and $T_h$, with controllable thermal valves $v_c$ and $v_h$, respectively. Depending on the stroke of the thermodynamic cycle, these valves switch between open and closed configurations, thereby coupling or decoupling the system from the corresponding thermal bath and thus regulating the heat exchange with it. Such controlled heat flow enables the realization of JDTM \cite{22marchegiani2016self,23pal2022josephson}. The loop incorporates a gate-tunable weak-link region in MATBG, driven into a non-superconducting, valley-polarized state at filling factor $\nu = -0.5$ (two holes per moire unit cell) \cite{37xie2023varphi}. The resulting valley polarization, together with trigonal warping of the MATBG Fermi surface, breaks time-reversal and intra-valley inversion symmetries, giving rise to the JDE observed in MATBG-JJs \cite{37xie2023varphi,38hu2023josephson}.

To model the junction, we consider an effective one-dimensional Hamiltonian corresponding to the $k_y = 0$ channel \cite{37xie2023varphi,38hu2023josephson}, given by $\mathcal{H}_{1D} = \frac{1}{2} \sum_{\eta,\gamma} \int dx \, \Psi_{\eta,\gamma}^\dagger(x)\, \hat{H}_{\eta,\gamma}(x)\, \Psi_{\eta,\gamma}(x)$, where $\eta = \pm$ denotes the valley index and $\gamma = \pm$ labels the right- and left-moving modes near the Fermi energy. The Nambu spinor is defined as $\Psi_{\eta,\gamma}(x) = \left(\psi_{\eta,\gamma}(x), \psi_{-\eta,-\gamma}^\dagger(x)\right)^T$, and the corresponding Hamiltonian is,

\begin{equation}
\hat{H}_{\eta,\gamma}(x) =
\begin{bmatrix}
-i\hbar v_{f,\eta,\gamma}(x)\partial_x + \eta \Delta_v(x) & \bm{\Delta}(x) \\
\bm{\Delta}(x) & -i\hbar v_{f,-\eta,-\gamma}(x)\partial_x + \eta \Delta_v(x)
\end{bmatrix},
\label{eq1:BdgMat}
\end{equation}

where the position-dependent Fermi velocity is $v_{f,\eta,\gamma}(x) = v_{s,\eta,\gamma} [\Theta(-x) + \Theta(x-d)] + v_{v,\eta,\gamma}\Theta(x)\Theta(d-x)$, with $v_{s,\eta,\gamma}$ and $v_{v,\eta,\gamma}$ denoting the Fermi velocities in the superconducting and weak-link regions. $\boldsymbol{\Delta}(x)= \Delta_s \left(e^{i\phi/2}\Theta(-x) + e^{-i\phi/2}\Theta(x-d)\right)$, with \(\Delta_s \) being the superconducting pairing potential, and \(\phi=\phi_L-\phi_R\) is the superconducting phase difference. \(\Delta_s \) depends on critical temperature \(\theta_c\) as, \(\Delta_s/\Delta_{s0}=\tanh{(1.74\sqrt{(\theta_c-T)/T})}\) \cite{37xie2023varphi}), where \(\Delta_{s0}\) is the superconducting pairing potential at absolute zero temperature. The valley polarization potential is $\Delta_v(x) = \Delta_v \Theta(x)\Theta(d-x)$ and \(d\) is the length of the weakly-linked valley-polarized junction (see, Fig. \ref{fig:fig1}(b)). Using the Blonder–Tinkham–Klapwijk (BTK) formalism, the Andreev bound state (ABS) energies $\epsilon_\eta$ can be self consistently obtained using the following expression,

\begin{equation}
\cos\left[2 \left(\beta - \frac{\epsilon_\eta - \eta \Delta_v}{E_T}\right)\right]
=
\cos\left(\phi + \frac{\epsilon_\eta - \eta \Delta_v}{\eta E_A}\right),
\label{eq2:ABS_cond_text}
\end{equation}

where $\beta = \cos^{-1}(\epsilon_\eta/\Delta_s)$, $E_T = \hbar v_T/d$ is the Thouless energy, and $E_A = \hbar v_A/d$ characterizes the energy scale associated with trigonal warping. By solving Eq.\ref{eq2:ABS_cond_text} numerically we obtain \(\epsilon_{\eta,n}\) where \(n=1,2,\dots\) are identified with ABS spectrum. The derivation of ABS energies (Eq.\ref{eq2:ABS_cond_text}), is given in Supplemental material (SM) (see, Sec. SM.2) \cite{SMsoufy2026supplementary}. For MATBG-JD, \(\phi=0\), while for MATBG-JJ, \(\phi\ne 0\). Energy scales \(E_T\) and \(E_A\) govern the intra-valley splitting as well as the associated phase shift in the ABS spectrum. Here, $v_T = 4/(v_{v,+,+}^{-1} + v_{v,+,-}^{-1}+v_{v,-,+}^{-1} + v_{v,-,-}^{-1})$ and $v_A = 2/(v_{v,+,+}^{-1} - v_{v,+,-}^{-1}+v_{v,-,+}^{-1} - v_{v,-,-}^{-1})$. \(v_{v,+,+}\) and \(v_{v,+,-}\) corresponds to Fermi velocity of intra-valley incoming and outgoing modes for the valley \(\eta=+\), and \(v_{v,-,+}\) and \(v_{v,-,-}\) for the valley \(\eta=-\). The setup considered here in Fig. \ref{fig:fig1} is in the ballistic limit and inter-valley back-scattering that couples the two valleys is neglected \cite{37xie2023varphi,38hu2023josephson}. The velocities \(v_T\) and \(v_A\) capture the slope and asymmetries of band structure, which then dictate the nature of ABS energies obtained. The trigonal-warped Fermi surface breaks intra-valley inversion symmetry (see, SM Sec. SM.1 \cite{SMsoufy2026supplementary} for more details) \cite{37xie2023varphi,38hu2023josephson,koshino2018maximally}, leading to $v_{v,\eta,+} \neq v_{v,\eta,-}$. For AABLG, the absence of trigonal warping implies $E_A \to \infty$, which correspondingly simplifies Eq.~\eqref{eq2:ABS_cond_text} as  \cite{IM6massatt2025defect,IM7hsu2010anomalous,IM8xie2023gate},

\begin{equation}
\cos\left[2 \left(\beta - \frac{\epsilon_\eta - \eta \Delta_v}{E_T}\right)\right]
=
\cos\left(\phi\right).
\label{eq2:ABS_aablg}
\end{equation}

The ABS energies, currents and diode efficiencies are plotted in the End Matter (EM) (Sec. \ref{EM.1}). Quantum thermodynamic cycles are generally analyzed within the Maxwell-Boltzmann statistics, since therein bath temperatures considered are of the order of $\sim 100\,\mathrm{K}$, in the present letter, we consider bath temperatures of the order of $\sim 0.1\,\mathrm{K}$, hence, the analysis here is carried out using Fermi-Dirac statistics. Thermodynamic quantities like Free energy (\(F(\Delta_v,T)\)), Entropy (\(S(\Delta_v,T)\)) and Internal energy (\(U(\Delta_v,T)\)) are then evaluated as follows \cite{24vischi2019thermodynamics,23pal2022josephson,37xie2023varphi,IM4vinjanampathy2016quantum},

\begin{figure}[!htbp]
    \centering
    \includegraphics[width=0.85\linewidth]{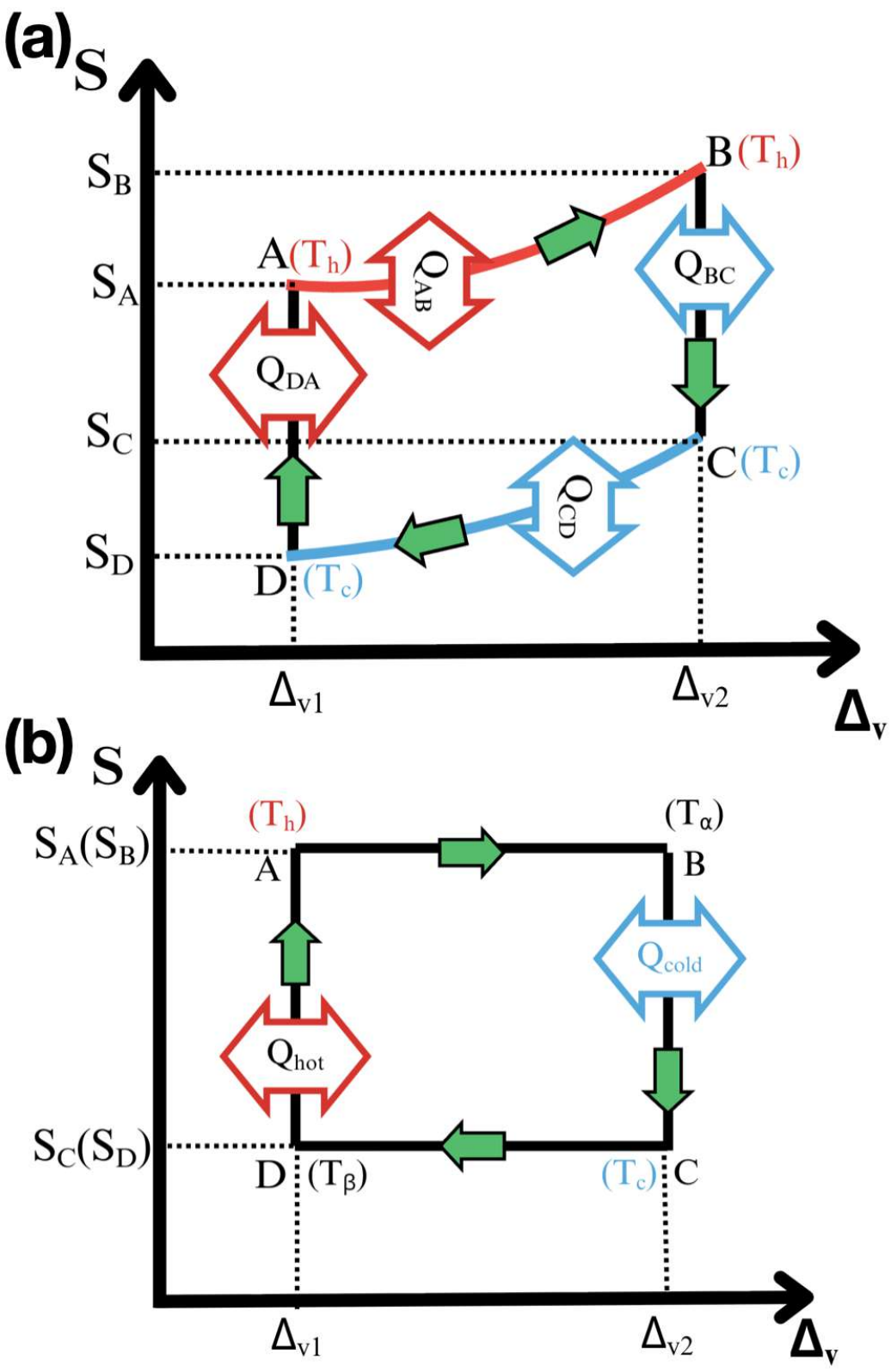}
    \caption{Entropy as a function of valley-polarization parameter $\Delta_v$ for (a) QSC, consisting of two quantum isothermal and two quantum isochoric strokes operating across temperatures \(T_c\) and \(T_h\), and (b) QOC, consisting of two quantum adiabatic and two quantum isochoric strokes operating across temperatures \(T_c,T_h,T_\alpha\) and \(T_\beta\). Red arrows indicate heat exchange with the hot reservoir, while blue arrows indicate heat exchange with the cold reservoir.}
    \label{fig:fig2}
\end{figure}

\vspace{-1em}

\begin{gather}
F(\Delta_v,T) = -k_B T \sum_{\eta,n} \ln\left(1+e^{-\frac{\epsilon_{\eta,n}(\Delta_v)}{k_BT}}\right), \nonumber\\
S(\Delta_v,T) = -\frac{\partial F}{\partial T}, \quad U(\Delta_v,T) = F + TS.
\end{gather}

Quantum thermodynamic cycles, constructed from a sequence of individual thermodynamic strokes, are defined through controlled variations of system parameters while selectively coupling the system to thermal baths to facilitate heat exchange. These cycles obey the first and second laws of thermodynamics \cite{IM2quan2007quantum,IM4vinjanampathy2016quantum}. In this letter, we focus on three standard four-stroke thermodynamic cycles, Quantum Stirling Cycle (QSC), Quantum Otto Cycle (QOC), and Quantum Carnot Cycle (QCC) \cite{IM2quan2007quantum,39soufy2025enhanced,IM4vinjanampathy2016quantum}. \textit{\textbf{For QCC, see EM (Sec. \ref{EM.2})}}.

\begin{table*}[!htbp]
\centering
\begin{tabular}{|c|c|c|c|c|c|}
\hline
\textbf{$Q_{\text{hot}}$} & \textbf{$Q_{\text{cold}}$} & \textbf{W} & \textbf{Operational Phase} & \textbf{Thermodynamic Constraints} & \textbf{Performance Metric} \\
\hline
$>$0 & $>$0 & $>$0 & \textcolor{red}{Forbidden} & Violates second law & -- \\
\hline
$>$0 & $>$0 & $<$0 & \textcolor{red}{Forbidden} & Violates first and second laws & -- \\
\hline
$>$0 & $<$0 & $>$0 & \textcolor{green}{Heat Engine} & $1 \le \beta \le \alpha$ & $\dfrac{\eta_{\text{HE}}}{\eta_c} = \dfrac{W_T}{Q_{\text{hot}}\,\eta_c}$ \\
\hline
$>$0 & $<$0 & $<$0 & \textcolor{green}{Cold Pump} & $0 \le \beta \le \min(1,\alpha)$ & $\text{COP}_{\text{CP}} = \left|\dfrac{Q_{\text{cold}}}{W_T}\right|$ \\
\hline
$<$0 & $>$0 & $>$0 & \textcolor{red}{Forbidden} & 
\begin{tabular}{@{}c@{}}
$\beta \le 1$ (First Law) \\
$\beta \ge \alpha \ge 1$ (Second Law)\\
Incompatible constraints 
\end{tabular} & -- \\
\hline
$<$0 & $>$0 & $<$0 & \textcolor{green}{Refrigerator} & $\beta \ge \max(1,\alpha)$ & $\dfrac{\text{COP}_{\text{RE}}}{\text{COP}_c} = \dfrac{Q_{\text{cold}}}{|W_T|\,\text{COP}_c}$ \\
\hline
$<$0 & $<$0 & $>$0 & \textcolor{red}{Forbidden} & Violates first law & -- \\
\hline
$<$0 & $<$0 & $<$0 & \textcolor{green}{Joule Pump} & $0 \le \beta < \infty$ & -- \\
\hline
\end{tabular}
\caption{Operational phases of quantum thermodynamic cycles, their thermodynamic constraints (obtained from first and second law), and the normalized performance metrics. Here, \(\alpha = T_h/T_c \ge1\), and \(\beta = |Q_{\text{hot}}/Q_{\text{cold}}|\). \(\eta_c=1-T_c/T_h\) is the efficiency of Carnot heat engine, \(\text{COP}_c=T_c/(T_h-T_c)\) is the coefficient of performance of Carnot refrigerator. }

\label{tableonly}
\end{table*}

\textbf{\textit{Quantum Stirling Cycle (QSC):}} The cycle comprises of four strokes, two isothermal and two isochoric (Fig.~\ref{fig:fig2}(a)) \cite{25scharf2020topological,23pal2022josephson,41soufy2026high}. It begins at $A$ with valley-polarization $\Delta_{v1}$. In the first stroke, the system undergoes an isothermal process at the hot bath temperature $T_h$ (with valve $v_h$ open), during which $\Delta_{v1}$ is brought to $\Delta_{v2}$ quasistatically, maintaining thermal equilibrium and exchanging heat $Q_{AB}=T_h(S_B-S_A)$. In the second stroke, valve $v_h$ is closed and $v_c$ is opened, leading to an isochoric process at fixed $\Delta_{v2}$, during which no work is done and heat exchange with the cold bath is $Q_{BC}=U_C-U_B$. The third stroke is an isothermal process at the cold bath temperature $T_c$, where the system is driven from $\Delta_{v2}$ back to $\Delta_{v1}$, with heat $Q_{CD}=T_c(S_D-S_C)$ exchanged with the cold bath. Finally, in the fourth stroke, valve $v_c$ is closed, and $v_h$ is reopened, allowing the system to thermalize isochorically at fixed $\Delta_{v1}$ and return to its initial state, with heat $Q_{DA}=U_A-U_D$ with the hot bath. The total heat exchanged with the hot bath and cold bath are $Q_{\text{hot}}=Q_{AB}+Q_{DA}$ and $Q_{\text{cold}}=Q_{BC}+Q_{CD}$, respectively.

\begin{figure*}[!htbp]
    \centering
    \includegraphics[width=0.9\linewidth]{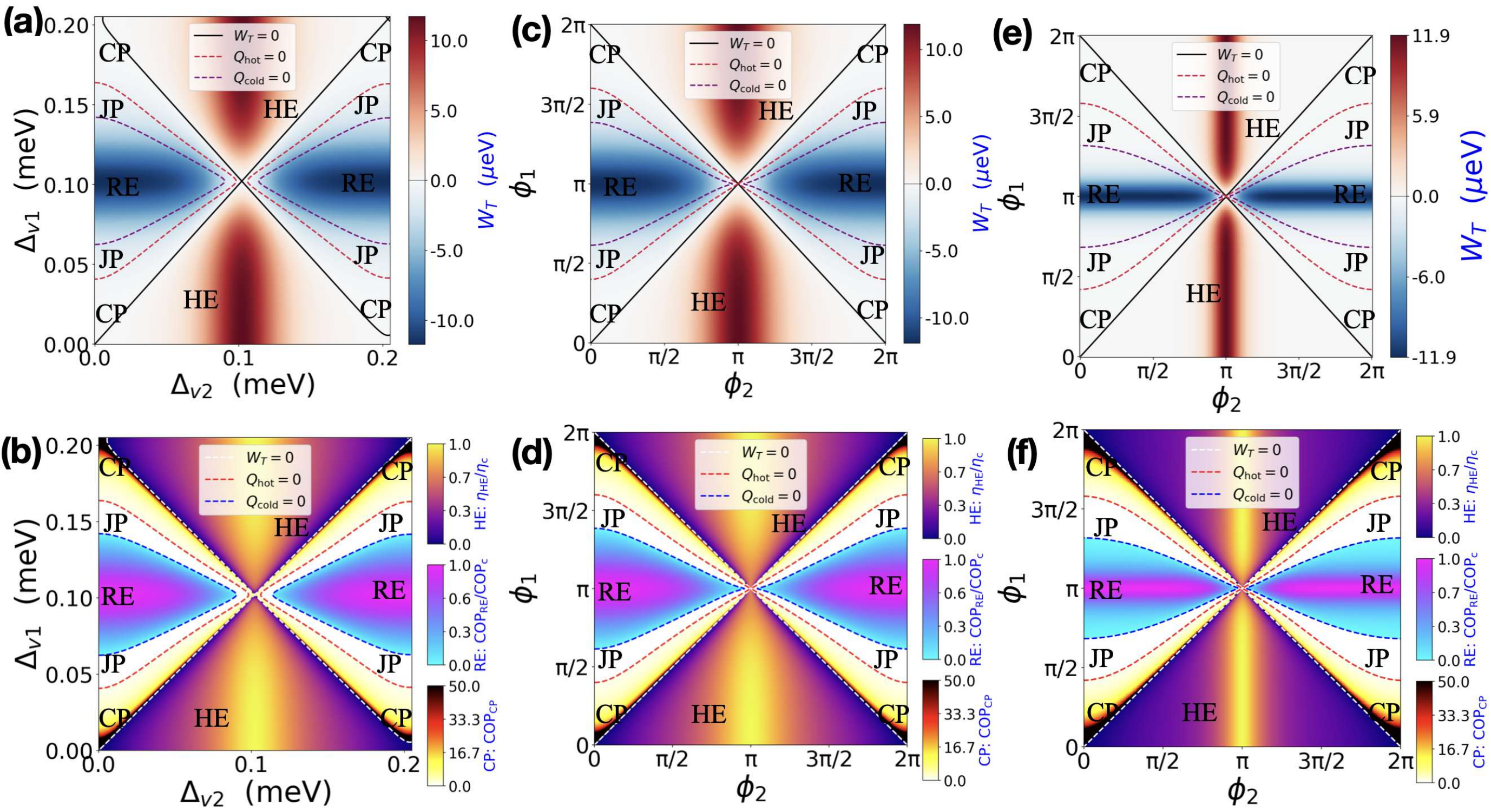}
    \caption{(a,b) Work and performance metrics of MATBG-JDTM exhibiting a QSC. (c,d) Work and performance metrics of MATBG-JJTM (e,f) for AABLG-JJTM for a QSC, in heat engine (HE), refrigerator (RE), cold pump (CP), and Joule pump (JP) regimes. The bath temperatures are $T_h=0.1\,\text{K}$ and $T_c=0.05\,\text{K}$.}
    \label{fig:fig3}
\end{figure*}

\textbf{\textit{Quantum Otto Cycle (QOC):}} The cycle consists of two adiabatic and two isochoric strokes (Fig.~\ref{fig:fig2}(b)) \cite{IM2quan2007quantum,39soufy2025enhanced}, starting at $A$ where the system is in equilibrium at temperature $T_h$ with valley-polarization $\Delta_{v1}$, and both valves closed to ensure isolation. In the first stroke, the system undergoes a slow adiabatic transformation in which $\Delta_{v1}$ changes to $\Delta_{v2}$; the entropy remains constant, leading to an effective temperature $T_\alpha$ defined by $S_A(\Delta_{v1},T_h)=S_B(\Delta_{v2},T_\alpha)$. In the second stroke, the valve $v_c$ is opened and the system thermalizes isochorically with the cold bath at temperature $T_c$ at fixed $\Delta_{v2}$, exchanging heat $Q_{\text{cold}}=U_C-U_B$. Next, the system is again isolated and driven adiabatically from $\Delta_{v2}$ to $\Delta_{v1}$, preserving entropy and reaching a second effective temperature $T_\beta$ such that $S_C(\Delta_{v2},T_c)=S_D(\Delta_{v1},T_\beta)$. In the final stroke, the valve $v_h$ is opened and the system thermalizes isochorically back to $T_h$, absorbing heat $Q_{\text{hot}}=U_A-U_D$ and completing the cycle. Unlike the QSC, the QOC involves four characteristic temperatures, $T_h$, $T_c$, $T_\alpha$, and $T_\beta$, all of which must remain below the superconducting critical temperature $\theta_c$ to preserve superconductivity. For MATBG, $\theta_c \approx 1.7\,\mathrm{K}$~\cite{26cao2018unconventional}, and we ensure that the bath temperatures lie well below this limit.

\textbf{\textcolor{red}{Operational Regimes and Performance analysis:}}  The total work performed over a complete cycle follows from the first law of thermodynamics and is given by $W_T = Q_{\text{hot}} + Q_{\text{cold}}$ and the second law imposes the constraint $Q_{\text{hot}}/T_h + Q_{\text{cold}}/T_c \le 0$ \cite{39soufy2025enhanced}. The operating mode of the cycle is determined by the relative signs of \(W_T\), \(Q_{\text{hot}}\), and \(Q_{\text{cold}}\), leading to distinct thermodynamic regimes including the heat engine (HE), refrigerator (RE), cold pump (CP), and Joule pump (JP). All remaining sign combinations are thermodynamically prohibited, and the performance of each allowed regime can be characterized through an appropriate figure of merit. Table~\ref{tableonly} summarizes all admissible operational phases along with their corresponding performance measures, where $\alpha = T_h/T_c\ge1$ and $\beta = |Q_{\text{hot}}/Q_{\text{cold}}|$ \cite{23pal2022josephson,39soufy2025enhanced}. 

For each cycle, we compare the performance of MATBG-JDTM, characterized by valley-polarization $\Delta_{v1}, \Delta_{v2}$ and with phase difference $\phi_1 = \phi_2 = 0$, versus MATBG-JJTM and AABLG-JJTM, where $\phi_1, \phi_2$ are non zero, while $\Delta_{v1} = \Delta_{v2} = 0$. More details on how a JJTM is operated via a superconducting phase in the quantum thermodynamic cycle is discussed in SM (Sec. SM.3) \cite{SMsoufy2026supplementary}.  Other parameters are $T_h = 0.1\,\text{K}$, $T_c = 0.05\,\text{K}$, and $\Delta_s = 0.1\,\text{meV}$. Codes used to obtain the results presented in this letter are available in \cite{Code}.

\begin{figure*}[!htbp]
\centering
\includegraphics[width=0.9\linewidth]{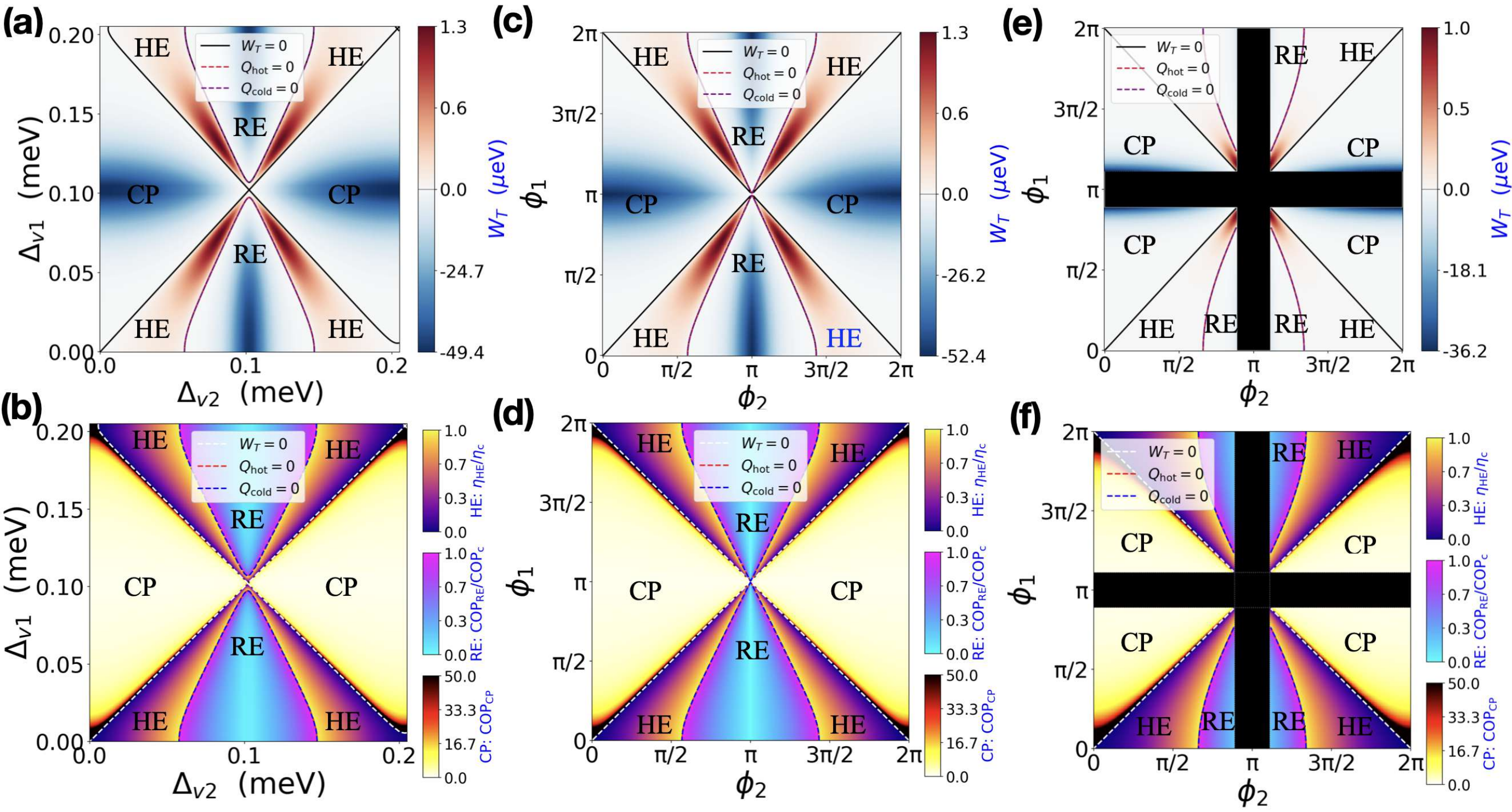}
\caption{(a,b) Work and performance metrics of MATBG-JDTM exhibiting a QOC. (c,d) Work and performance metrics of MATBG-JJTM (e,f) for AABLG-JJTM for a QOC, in the heat engine (HE), refrigerator (RE), and cold pump (CP) regimes. The bath temperatures are $T_h=0.1\,\text{K}$ and $T_c=0.05\,\text{K}$.}
\label{fig:fig4new}
\end{figure*}

\begin{figure*}[!htbp]
    \centering
    \includegraphics[width=0.9\linewidth]{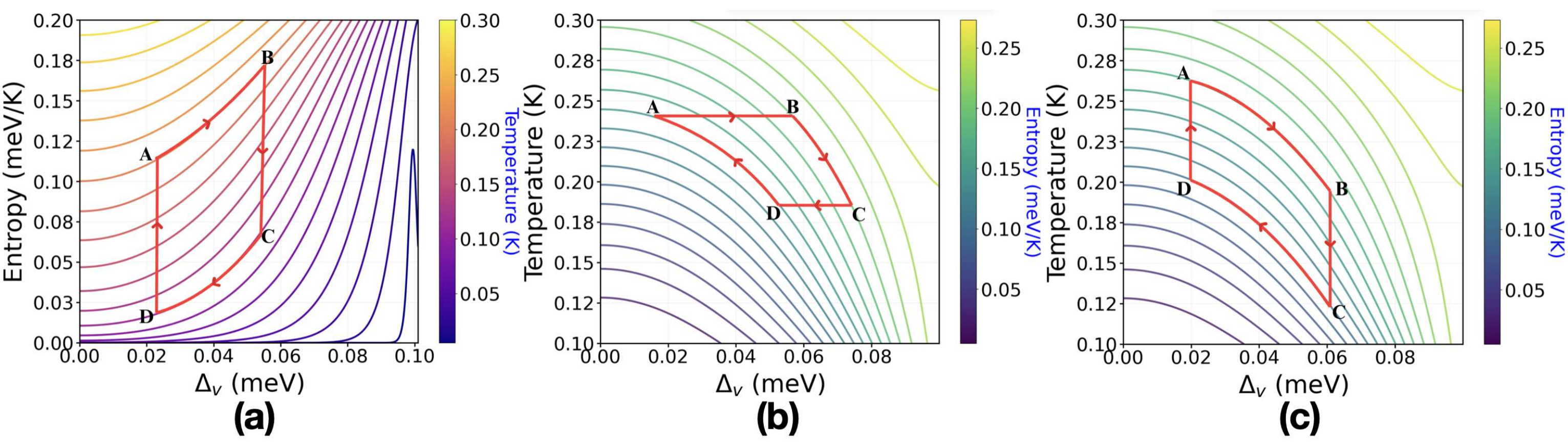}
    \caption{(a) Isothermal contour plots depicting QSC in MATB-JDTM, (b),(c) are isoentropic contours for MATBG-JDTM depicting QOC and QCC, respectively.}
    \label{fig:fig6}
\end{figure*}

Fig.~\ref{fig:fig3}(a,b) show the operational regimes and performance of MATBG-JDTM, Fig.~\ref{fig:fig3}(c,d) present the corresponding results for MATBG-JJTM, while Fig.~\ref{fig:fig3}(e,f) corresponds to AABLG-JJTM, all for QSC. MATBG-based devices exhibit very similar qualitative and quantitative features regardless of whether the valley-polarization potential $\Delta_v$ (JDTM) or the superconducting phase difference $\phi$ (JJTM) is the control parameter. Both MATBG-JDTM and MATBG-JJTM display high-performance regions approaching the Carnot limit, along with sharp cold-pump enhancements near operational phase boundaries (i.e, where $W_T =0$). AABLG-JJTM (Fig.~\ref{fig:fig3}(e,f)), on the other hand, shows a strongly suppressed work output over much of the parameter space, with significant performance enhancements only near $\phi_{1,2}\approx\pi$. Valley polarization in MATBG-JJs provides a more experimentally accessible control knob than superconducting phase tuning, suggesting a more practical route to thermodynamic operation than conventional JJ thermal machines. Moreover, the presence of trigonal warping, together with the reduced Fermi velocity in MATBG, significantly broadens the high-efficiency operating regime, as seen when compared with AABLG-JJTM. The underlying reason for the similar behavior observed in MATBG-JDTM and MATBG-JJTM versus AABLG-JJTM can be understood from the ABS spectrum, the anomalous currents, the occupation number per energy levels, and the corresponding entropy characteristics, which are discussed in detail in SM (Sec. SM.4) \cite{SMsoufy2026supplementary} and EM.\ref{EM.1}.

Fig.~\ref{fig:fig4new}(a,b) depict the operational regimes and performance of MATBG-JDTM, while Fig.~\ref{fig:fig4new}(c,d) show the corresponding results for MATBG-JJTM operating as a QOC. Similar to the QSC case, both configurations exhibit nearly identical operational phases and performance. The performance of heat engine and refrigeration modes peak near the operational phase boundaries; however, these regions also correspond to vanishing work output ($W_T = 0$). Compared to the QSC, the net work output in the heat engine regime is significantly reduced for QOC, while regions requiring larger work input for refrigeration exhibits lower performance, highlighting a trade-off between work and efficiency \cite{IM2quan2007quantum}. The cold pump coefficient of performance similarly shows a sharp enhancement near a phase transition, where $W_T =0 $.  Fig.~\ref{fig:fig4new}(e,f) shows the operational regime of AABLG-JJTM operating as a QOC. Similar to the QSC, most of the parameter space yields zero work output. Near $\phi_{1,2}\approx\pi$, the ABS energies approach zero, causing the entropy to diverge, and the entropy-matching condition (see, Eqs. (SM.17) and (SM.18) in \cite{SMsoufy2026supplementary}) requires large relaxation temperatures $T_\alpha$ and $T_\beta$, effectively pushing the system out of the superconducting regime. This region is therefore excluded from performance analysis (detailed explanation provided in the SM (Sec. SM.5) \cite{SMsoufy2026supplementary}). MATBG-based JJ thermal machines (both JJTM and JDTM) remain well within the superconducting regime across the full QOC parameter space, with relaxation temperatures consistently below critical temperatures, highlighting their superior thermodynamic stability. Thus, AABLG-JJTM does not act as a thermal machine for a wide range of parameters, which is a limitation compared to MATBG-JDTM and MATBG-JJTM which are universal.

\textbf{\textcolor{red}{Experimental Realization and Conclusion:}} Fig.~\ref{fig:fig6} illustrates how the QSC, QOC, and QCC can be implemented in MATBG-JDTM through controlled tuning of the valley-polarizing potential and temperature. These cycles can be realized experimentally by executing the constituent thermodynamic strokes sequentially. Thermal valves have also been experimentally realized using quantum point contacts in a two-dimensional (2D) electron gas \cite{IM15amado2014ballistic}.
Heat exchange during quantum isochoric strokes is achieved by coupling the junction to baths while keeping the system parameters fixed, whereas quantum isothermal strokes are implemented by varying the system parameters while maintaining thermal equilibrium with the bath; both processes require the corresponding thermal valves to remain open. Adiabatic strokes, on the other hand, involve no heat exchange, with the work directly determined by the change in internal energy; these are performed with both valves closed and under slow driving conditions to preserve quasi-equilibrium and a well-defined temperature. In this letter, we showed that MATBG-JDTM exhibits identical operational regimes and performance as phase-controlled MATBG-JJTM. A comparison with AABLG-JJTM highlights the advantages of MATBG, where trigonal warping and reduced Fermi velocity lead to wider regions of finite work output and high performance across all thermodynamic cycles considered. Conventional phase tuning in Josephson devices generally relies on magnetic-flux control, which can introduce flux-noise-induced decoherence and cross-talk between neighboring elements \cite{IM12bialczak20071,IM13van2004decoherence}. Electrostatic gating, which tunes the valley-polarization, offers a complementary route with local tunability and reduced dependence on external magnetic fields \cite{37xie2023varphi,IM14de2021gate}. Recent advances in gate-defined Josephson diode devices in MATBG \cite{35rothstein2026gate,32de2021gate}, along with demonstrations of superconducting thermal machines \cite{IM9uusnakki2026initial,IM10araujo2024superconducting,IM11aamir2025thermally}, provide a promising route toward realizing MATBG-JDTM.

\bibliographystyle{unsrt} % Shows titles in references
\bibliography{ref}

\clearpage

\onecolumngrid

\section*{End Matter}
Details of the Andreev bound states, current--phase relations, and diode efficiency for different junction parameters are presented in Sec. \ref{EM.1}, while Sec. \ref{EM.2} discusses the performance and results of MATBG-JDTM, MATBG-JJTM, and AABLG-JJTM operating under the Quantum Carnot Cycle (QCC).

\setcounter{subsection}{0} % Reset subsection counter
\renewcommand{\thesubsection}{EM.\arabic{subsection}} % Set numbering to B.1, B.2, etc

\begin{figure}[!htbp]
    \centering
    \includegraphics[width=0.87\textwidth]{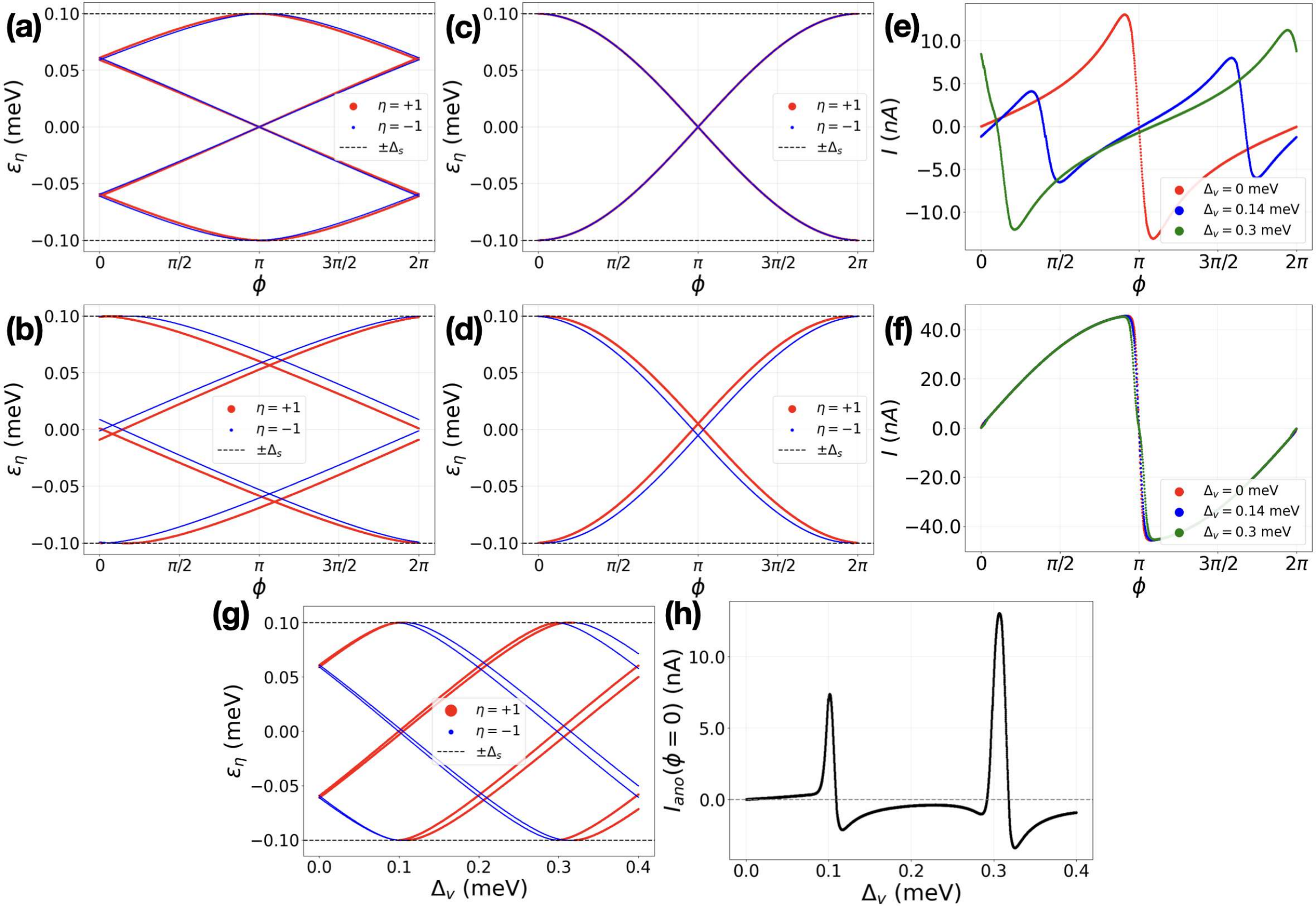}
    \caption{(a,b) ABS spectra of MATBG-JJs at $\Delta_v = 0$ and $0.3\,\text{meV}$. (c,d) ABS spectra of AABLG-JJs at $\Delta_v = 0$ and $0.3\,\mathrm{meV}$. Corresponding current-phase relations (CPR) for (e) MATBG-JJs and (f) AABLG-JJs. (g,h) ABS spectra and corresponding anomalous current $I_{\text{ano}}$ as a function of $\Delta_v$ at $\phi=0$, the currents are calculated at \(T = 0.02\, \mathrm{K}\). The length of the weak-lined junction \(d = 80\  \mathrm{nm}\) (see, Ref.\cite{37xie2023varphi}).}
    \label{fig:NEWfig7}
\end{figure}

\twocolumngrid

\subsection{Andreev Bound State, Current Phase Relation and Josephson Diode efficiency in MATBG-JJs}
\label{EM.1}

\begin{figure}[!htbp]
    \centering
  \includegraphics[width=0.85\linewidth]{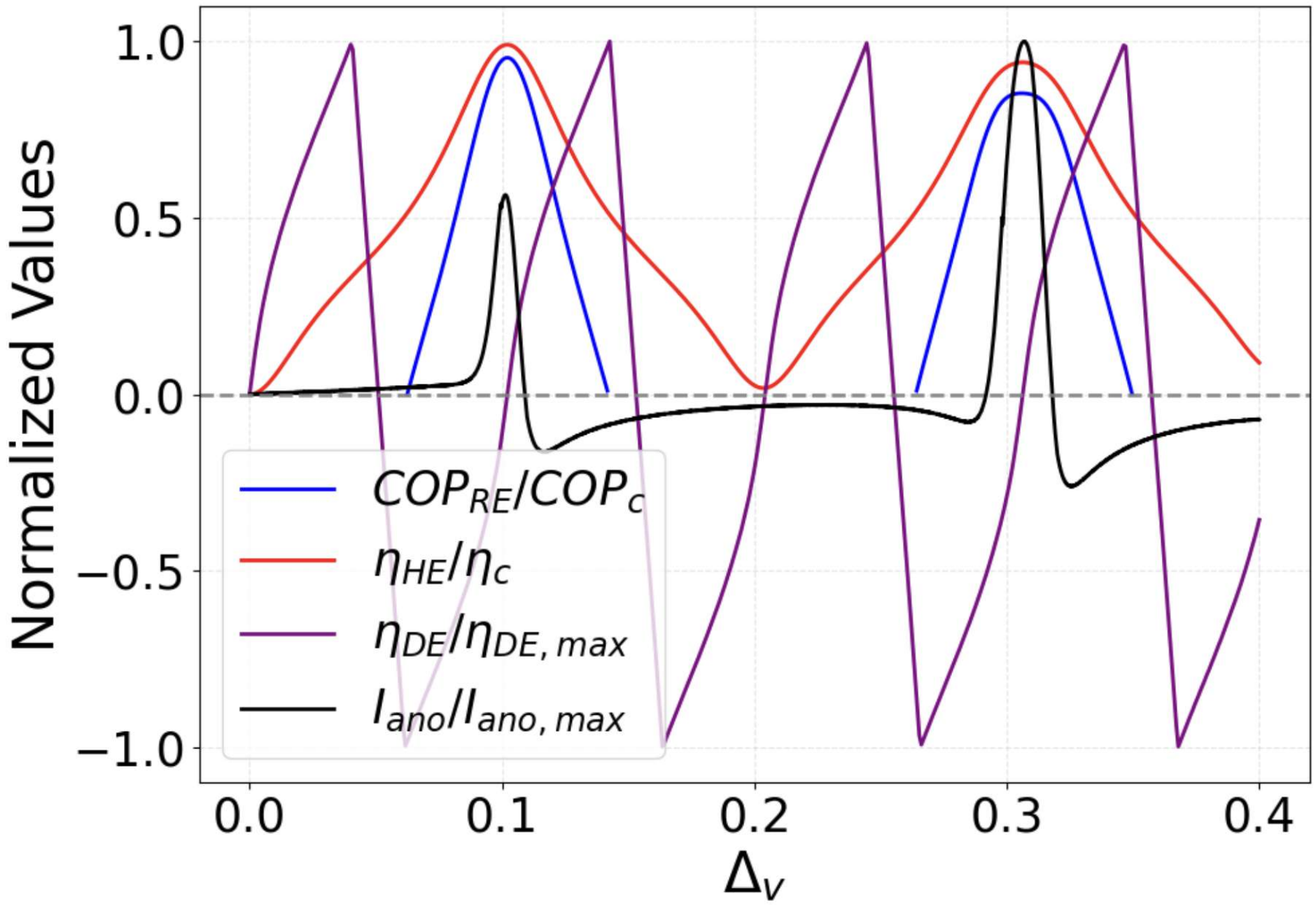}
    \caption{\(\frac{\eta_{DE}}{\eta_{DE,\max}}\), \(\frac{I_{\mathrm{ano}}}{I_{\mathrm{ano},\max}}\), and the QSC performance metrics \(\frac{\eta_{HE}}{\eta_c}\) and \(\frac{\mathrm{COP}_{RE}}{\mathrm{COP}_c}\), plotted as functions of \(\Delta_v\), where \(\frac{\eta_{HE}}{\eta_c}\) is evaluated for \(\Delta_v=\Delta_{v2}\) with \(\Delta_{v1}=0\), and \(\frac{\mathrm{COP}_{RE}}{\mathrm{COP}_c}\) for \(\Delta_v=\Delta_{v1}\) with \(\Delta_{v2}=0\).}
    \label{fig:fig8new}
\end{figure}

Figs.~\ref{fig:NEWfig7}(a,b) show the Andreev bound state (ABS) spectra for MATBG-based JJ at valley-polarization values $\Delta_v = 0 \ \text{meV}$ and $\Delta_v = 0.3\ \text{meV}$, respectively, obtained from Eq.~\eqref{eq2:ABS_cond_text}. The energy scales are chosen as $E_T = 0.065\,\text{meV}$ and $E_A = 1.2\,\text{meV}$, following Refs.~\cite{37xie2023varphi,38hu2023josephson}; these parameters are treated phenomenologically to capture the essential features of MATBG junctions with valley polarization ~\cite{xie_private_comm}. Figs.~\ref{fig:NEWfig7}(c,d) show the corresponding ABS spectra for AABLG-based junctions at the same $\Delta_v$ values. In AABLG, the absence of trigonal warping implies $v_{v,\eta,+} = v_{v,\eta,-}$, see below Eq.\eqref{eq2:ABS_cond_text}, in main, leading to $E_A \to \infty$, while the larger Fermi velocity ($\sim 10^5\text{m/s}$) yields $E_T = 5.75 \,\text{meV}$ \cite{IM8xie2023gate}. Consequently, the valley-induced splitting with increasing $\Delta_v$ is significantly weaker compared to MATBG. The current-phase relation (CPR), from the free energy is

\begin{equation}
I(\phi,\Delta_v,T) = \frac{2e}{\hbar}\frac{\partial F(\phi,\Delta_v,T)}{\partial\phi}.
\end{equation}

\vspace{-1em}

\begin{figure}[!htbp]
    \centering
    \includegraphics[width=0.8\linewidth]{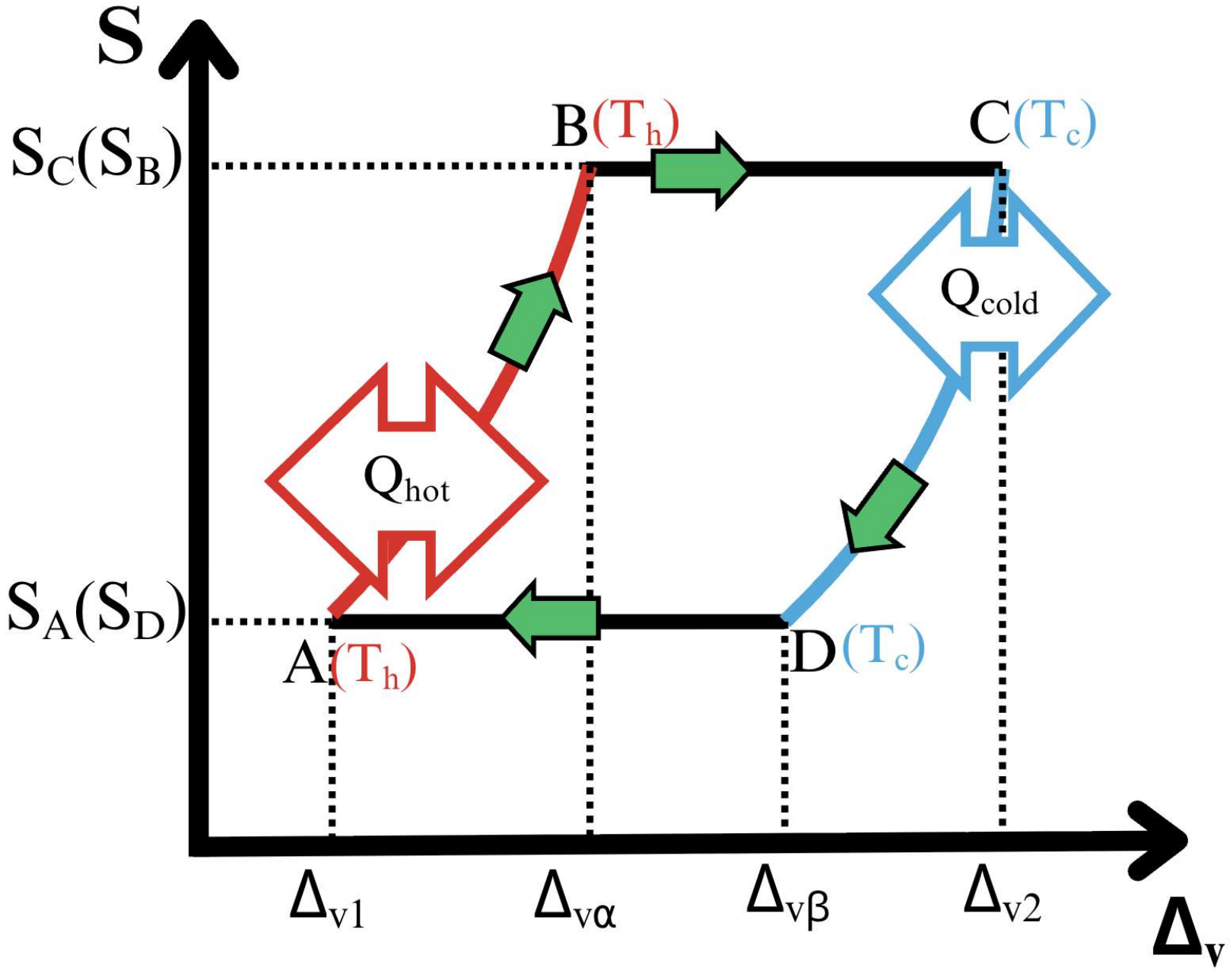}
    \caption{Entropy as a function of $\Delta_v$ for QCC in MATBG-JDTM, consisting of two quantum isothermal and two quantum adiabatic strokes.}
    \label{fig:fig9_newnew}
\end{figure}

The corresponding CPR for both systems is shown in Fig.\ref{fig:NEWfig7}(e) for MATBG and Fig.\ref{fig:NEWfig7}(f) for AABLG. In MATBG junctions, increasing $\Delta_v$ induces a finite super-current at $\phi=0$, indicating $\phi_0$-junction behavior, whereas AABLG-JJs show only marginal changes in the CPR and do not exhibit anomalous phase shifts. Figs.~\ref{fig:NEWfig7}(g,h) further present the ABS spectra together with the corresponding anomalous current $I_{\mathrm{ano}}$ which is the current evaluated at $\phi=0$ as a function of $\Delta_v$ for MATBG-JD. The valley-induced splitting increases with $\Delta_v$, with pronounced peaks in $I_{\mathrm{ano}}$ appearing when the ABS energies approach the superconducting gap.The diode rectification efficiency is given as, $ \eta_{\mathrm{DE}} = \frac{|I_{\max}| - |I_{\min}|}{|I_{\max}| + |I_{\min}|},$ where, $I_{\max} = \max_{\phi} I(\phi,\Delta_v,T)\  \text{and}\ 
I_{\min} = \min_{\phi} I(\phi,\Delta_v,T)$. Fig.~\ref{fig:fig8new} depicts the normalized values \(\frac{\eta_{DE}}{\eta_{DE,\max}}\), \(\frac{I_{\mathrm{ano}}}{I_{\mathrm{ano},\max}}\), \(\frac{\eta_{HE}}{\eta_c}\), and \(\frac{\mathrm{COP}_{RE}}{\mathrm{COP}_c}\) as functions of \(\Delta_v\). The quantities \(\frac{\eta_{HE}}{\eta_c}\) and \(\frac{\mathrm{COP}_{RE}}{\mathrm{COP}_c}\) are evaluated for the QSC at \(T_c=0.05\,\mathrm{K}\) and \(T_h=0.1\,\mathrm{K}\), where the heat engine efficiency is obtained as a function of \(\Delta_v=\Delta_{v2}\) with \(\Delta_{v1}=0\), while the refrigeration COP is obtained as a function of \(\Delta_v=\Delta_{v1}\) with \(\Delta_{v2}=0\) (see, Fig.\ref{fig:fig3}(b) in the main text). The quantities \(\frac{\eta_{DE}}{\eta_{DE,\max}}\) and \(\frac{I_{\mathrm{ano}}}{I_{\mathrm{ano},\max}}\) are calculated at \(T=0.02\,\mathrm{K}\), with \(\eta_{DE,\max}\) and \(I_{\mathrm{ano},\max}\) denoting the respective maxima within the range considered. The performance of the MATBG-JDTM peaks near the maximum anomalous current and exhibits a sharp variation around this region.

\subsection{Quantum Carnot Cycle}
\label{EM.2}

\begin{widetext}

\begin{figure*}[!b]
\centering
\includegraphics[width=0.85\linewidth]{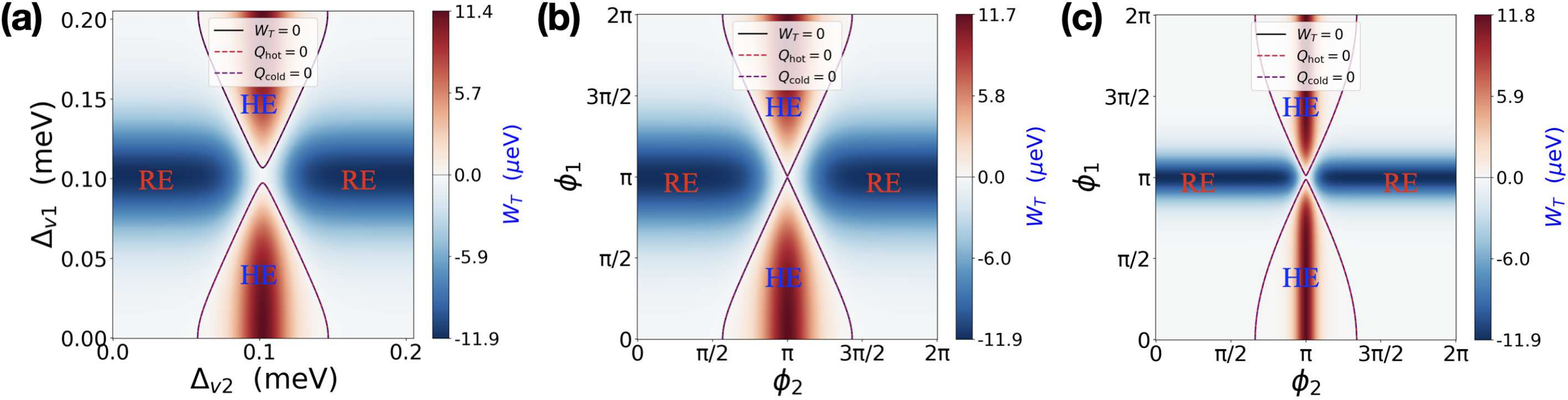}
\caption{Operational regimes and work exchanged for (a) MATBG-JDTM and (b) MATBG-JJTM and (c) AABLG-JJTM under QCC in their respective parameter space. The operational phases include heat engine (HE), and refrigerator (RE). The temperatures are fixed at $T_h=0.1\,\text{K}$ and $T_c=0.05\,\text{K}$.}
    \label{fig:fig10}
\end{figure*}
    
\end{widetext}

The cycle consists of two isothermal and two adiabatic strokes, as shown in Fig.~\ref{fig:fig9_newnew}. From $\Delta_{v1}$ at $T_h$, an isothermal expansion with $v_h$ open reaches $\Delta_{v\alpha}$ and absorbs heat $Q_{\text{hot}}=T_h(S_B-S_A)$. An adiabatic stroke with both valves closed then brings the system to $\Delta_{v2}$ at $T_c$, satisfying $S_B\left(\Delta_{v\alpha},T_h\right)=S_C\left(\Delta_{v2},T_c\right)$. Next, an isothermal compression at $T_c$ with $v_c$ open reaches $\Delta_{v\beta}$ and releases heat $Q_{\text{cold}}=T_c(S_D-S_C)$. Finally, an adiabatic stroke with both valves closed returns the system to $\Delta_{v1}$ at $T_h$, satisfying $S_D\left(\Delta_{v\beta},T_c\right)=S_A\left(\Delta_{v1},T_h\right)$. Fig.~\ref{fig:fig10}(a,b) depict the operational regimes of MATBG-JDTM and MATBG-JJTM, respectively, under a QCC. In both cases, the cycle is reversible and thus supports only heat engine (HE) and refrigerator (RE) modes \cite{39soufy2025enhanced}, operating at the Carnot efficiency $\eta_c = 1 - \frac{T_c}{T_h}$ in the HE regime and the Carnot coefficient of performance $COP_c = \frac{T_c}{T_h - T_c}$ in the RE regime. As in the previous cases, the results for MATBG-JDTM and MATBG-JJTM are nearly identical, indicating that the valley-polarizing gate potential can effectively serve as a control parameter analogous to the superconducting phase. Fig~\ref{fig:fig10}(c) show the corresponding operational regimes for AABLG-JJTM under a QCC. Consistent with earlier observations,  most of the parameter space corresponds to negligible work exchange, except $\phi_{1,2} \approx \pi$.

\clearpage

\onecolumngrid

\section*{\uline{Supplemental material} for "MATBG Josephson diode as an universal thermal machine"}

\begin{center}
    Hadi Mohammed Soufy\(^{1,2}\),  and Colin Benjamin\(^{1,2}\)

    \vspace{0.2em}

    \small{\textit{\(^1\)School of Physical Sciences, National Institute of Science Education and Research, HBNI, Jatni-752050, India}}

    \small{\textit{\(^2\)Homi Bhabha National Institute, Training School Complex, Anushakti Nagar, Mumbai, 400094, India}}
\end{center}

In the Supplemental Material (SM), we provide detailed theoretical and computational support for the results presented in the main text. In Sec.~\ref{SM.1}, we present a comprehensive description of the continuum model for twisted bilayer graphene (TBG), highlighting the emergence of trigonal warping and its absence in AA-stacked bilayer graphene (AABLG) as the twist angle approaches zero. We further demonstrate the resulting anisotropic Fermi velocity in MATBG, which is essential for realizing the diode effect. In Sec.~\ref{SM.2}, we provide detailed derivation of the Andreev bound state (ABS) condition from the effective 1D model for MATBG-based Josephson junctions with a valley-polarizing gate voltage in the weak-link region, as introduced in the main text, using the Blonder--Tinkham--Klapwijk (BTK) formalism following Ref.\cite{37xie2023varphi,38hu2023josephson}. In Sec.~\ref{SM.3}, we extend the framework to thermal machines based on TBG junctions, where both the valley-polarizing potential and superconducting phase act as control parameters; the diode-based thermal machine is obtained by setting the superconducting phase to zero, while the Josephson-based thermal machine is obtained by setting the valley polarization to zero. In Sec.~\ref{SM.4}, we analyze the Fermi--Dirac distribution and entropy for MATBG-JDTM, MATBG-JJTM, and AABLG-JJTM systems, providing additional insight into the underlying mechanisms governing the operational performance discussed in the main text. Sec.~\ref{SM.5}, provides a detailed analysis of the adiabatic condition in the Otto cycle, emphasizing that the relaxation temperature must remain below the critical temperature, a condition that fails in certain regions of AABLG parameter space, thereby restricting its operational regime. This condition is absent in Carnot cycles, since the isothermal strokes in between the adiabatic processes constrain the system to remain at the reservoir temperatures \(T_c
\) and \(T_h\) throughout the cycle. Finally, in Sec.~\ref{SM.6}, we outline the numerical procedure used to generate the results in the main text, starting from solving the self-consistent ABS equations to evaluating the performance of the thermal machines.

\setcounter{equation}{0}
\setcounter{figure}{0}
\renewcommand{\thefigure}{SM\arabic{figure}}
\renewcommand{\theequation}{SM\arabic{equation}}
\setcounter{subsection}{0} % Reset subsection counter
\renewcommand{\thesubsection}{SM.\arabic{subsection}} % Set numbering to B.1, B.2, etc

\subsection{Trigonal Warping effects in MATBG}
\label{SM.1}

\begin{wrapfigure}{r}{0.45\textwidth}
    \centering
    \includegraphics[width=\linewidth]{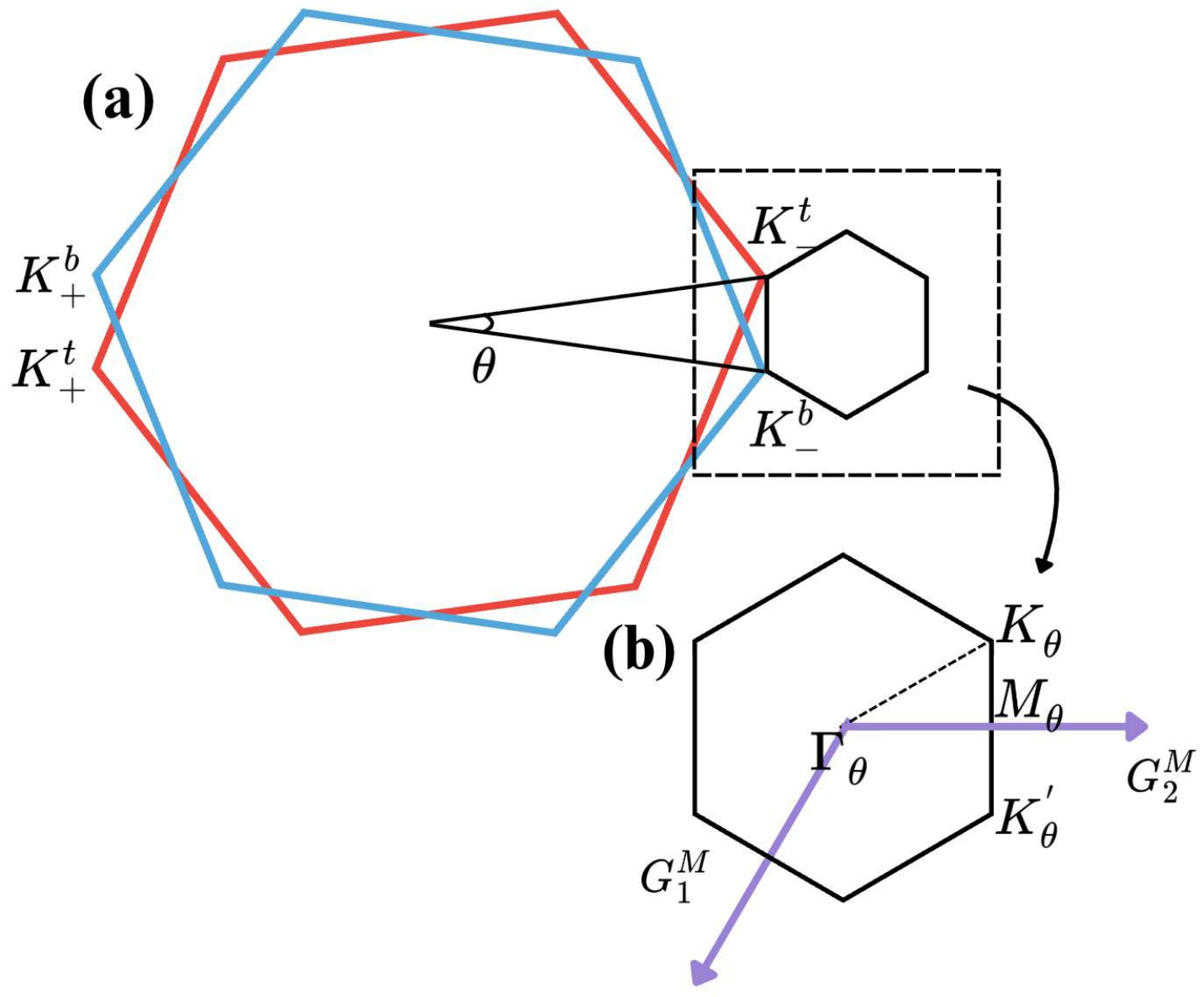}
    \caption{(a) Hexagonal Brillouin zones of the individual graphene layers rotated relative to each other by an angle $\theta$.  (b) Moiré Brillouin zone of twisted bilayer graphene, showing the high-symmetry points and the moiré reciprocal lattice vectors.}
    \label{Fig.SM01}
\end{wrapfigure}

We construct the atomic structure of twisted bilayer graphene (TBG) by starting from an AA-stacked bilayer configuration \cite{37xie2023varphi,koshino2018maximally}, in which the two graphene sheets are perfectly aligned such that their honeycomb lattices coincide. A relative twist is introduced between the layers by rotating layer $l=b$ (bottom) and $l=t$ (top) each by angles $-\theta/2$ and $+\theta/2$, respectively (see, Fig.\ref{Fig.SM01}). For a monolayer, the primitive lattice vectors are defined as 
$\mathbf{a}_1 = a(1,0)$ and $\mathbf{a}_2 = a(1/2, \sqrt{3}/2)$, where $a \approx 0.246\,\mathrm{nm}$ is the lattice constant of graphene \cite{37xie2023varphi,koshino2018maximally,29bistritzer2011moire}. The corresponding reciprocal lattice vectors are defined as 
$\mathbf{b}_1 = \frac{2\pi}{a}(1, -1/\sqrt{3})$ and 
$\mathbf{b}_2 = \frac{2\pi}{a}(0, 2/\sqrt{3})$. The rotation matrix is given by
\[
\mathcal{R}(\theta) =
\begin{pmatrix}
\cos\theta & -\sin\theta \\
\sin\theta & \cos\theta
\end{pmatrix}.
\]

Upon twisting, the lattice vectors for each layer transform as $\mathbf{a}_i^{l} = \mathcal{R}(\mp \theta/2)\,\mathbf{a}_i,$
where the upper (lower) sign corresponds to layer $l=b$ ($l=t$). The reciprocal lattice vectors transform analogously as $\mathbf{b}_i^{l} = \mathcal{R}(\mp \theta/2)\,\mathbf{b}_i.$ The reciprocal lattice vectors associated with the moiré superlattice are defined as,

\begin{equation}
\mathbf{G}_j^{M} = \mathbf{b}_j^{b} - \mathbf{b}_j^{t}=\frac{4\pi}{\sqrt{3}L_M}
\left(\cos\frac{2(j+1)\pi}{3}, \sin\frac{2(j+1)\pi}{3}\right),
\label{eqSM1}
\end{equation}

\begin{figure}[!htbp]
    \centering
    \includegraphics[width=0.85\linewidth]{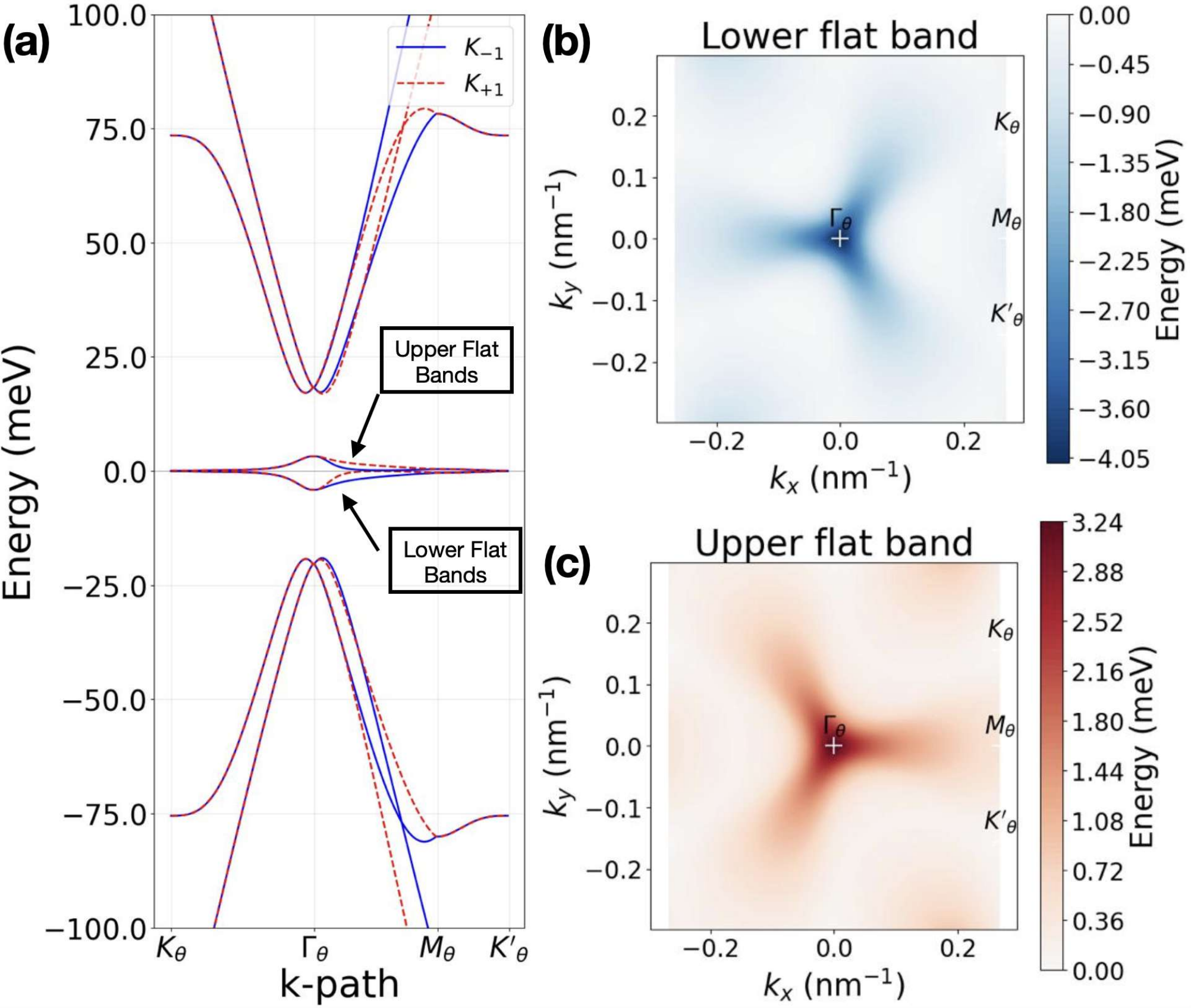}
    \caption{(a) Band structure of MATBG along the high-symmetry directions of the moiré Brillouin zone for both valleys, \(K_{+1}\) (dashed red) and \(K_{-1}\) (solid blue). Panels (b) and (c) show the trigonal warping of the lower and upper flat bands, respectively, for the \(\eta=+1\) valley.}
    \label{matbgband}
\end{figure}

where $j\in\{1,2\}$ and \(L_M=a/(2\sin{(\theta/2))} \) is the magnitude of moiré lattice vector \cite{koshino2018maximally}. In momentum space, this leads to a folding of the Brillouin zones, where the large hexagons correspond to the Brillouin zones of the individual layers, and the smaller hexagon represents the moiré Brillouin zone of TBG. The Dirac points of graphene are located at $
\mathbf{K}_{\eta}^{l} = -\eta \frac{2\mathbf{b}_1^{l} + \mathbf{b}_2^{l}}{3}$ \cite{koshino2018maximally}, where $\eta = \pm 1 $ labels the valley degree of freedom. The high-symmetry points of the moiré Brillouin zone are denoted as 
$\Gamma_\theta$, $M_\theta$, $K_\theta$, and $K'_\theta$. These points are crucial because they capture the essential symmetry and low-energy electronic properties of the moiré superlattice. Trigonal warping effects in MATBG are captured within the effective continuum model written in the sublattice basis, described by the Hamiltonian

\begin{figure}[!htbp]
    \centering
    \includegraphics[width=0.9\linewidth]{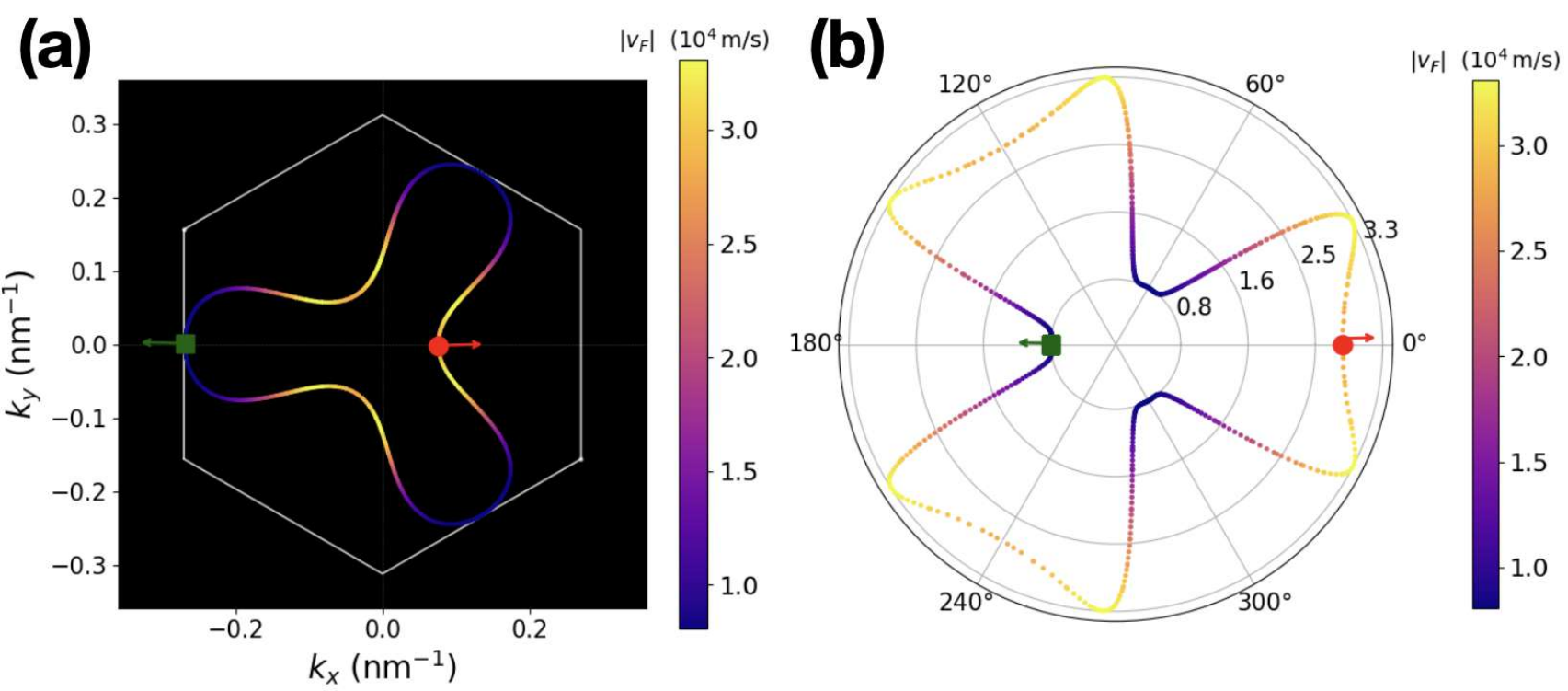}
    \caption{(a) Fermi surface at $\mu = 0.415\,\text{meV}$ with corresponding Fermi velocity distribution shown as a heat map. (b) Polar plot of the Fermi velocity illustrating its anisotropy for valley \(\eta=+1\). The dark green square and red circle corresponds to incoming/outgoing mode along the \(k_y=0\) channel which are not equal.}
    \label{trigwarp}
\end{figure}

\begin{figure}[!htbp]
    \centering
    \includegraphics[width=0.85\linewidth]{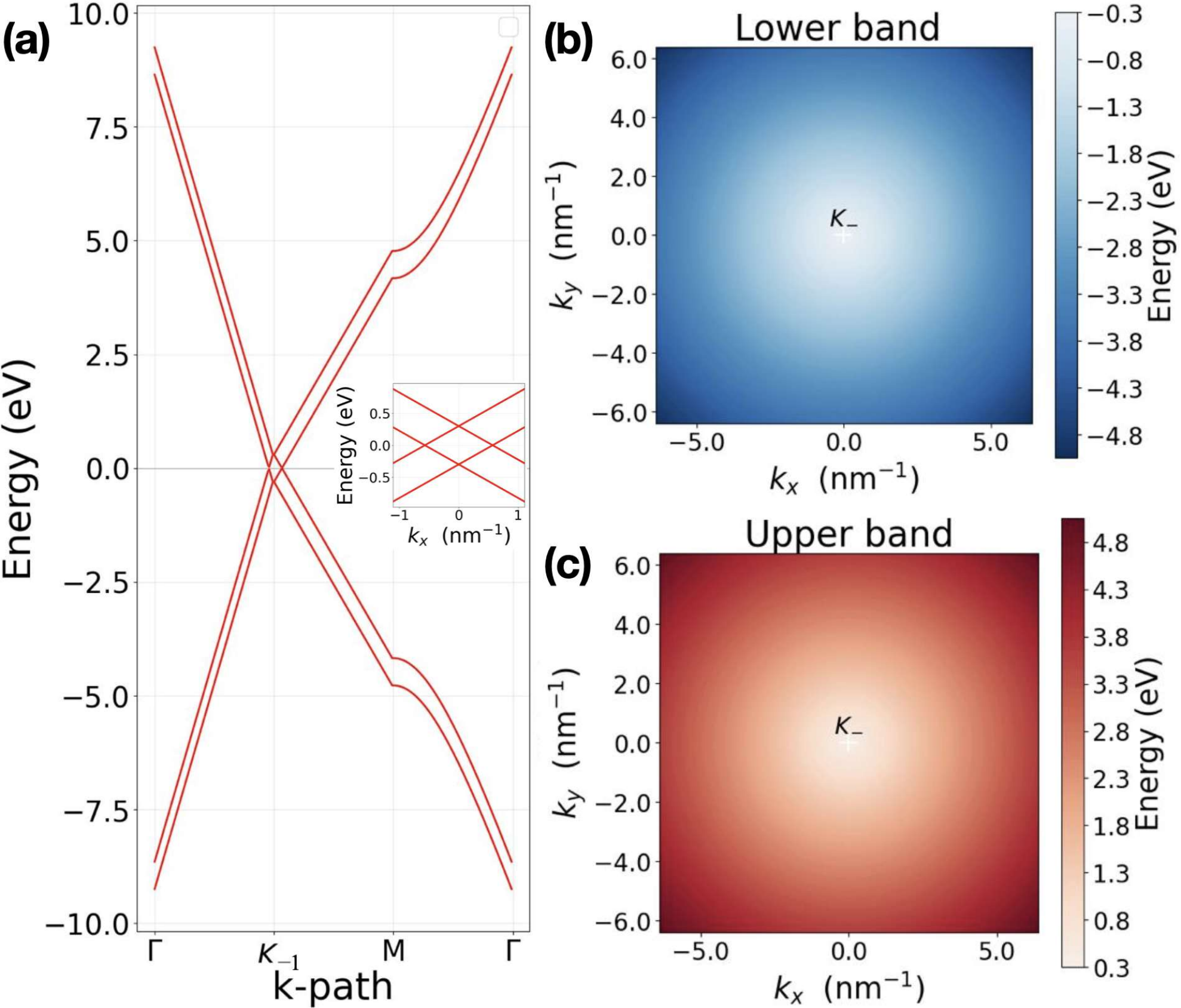}
    \caption{(a) Band structure of AABLG along the high-symmetry directions of the Brillouin zone for the \(K_{-1}\) valley. The inset shows the dispersion along \(k_x\) for \(k_y=0\). Panels (b) and (c) present contour plots of the lower and upper bands, respectively, illustrating the absence of trigonal warping.}
    \label{aablgband}
\end{figure}

\begin{equation}
\mathcal{H}_\eta(\mathbf{r}) =
\begin{bmatrix}
H^t_\eta & T \\
T^\dagger & H^b_\eta
\end{bmatrix},
\label{eqSM2}
\end{equation}
where the layer-resolved Dirac Hamiltonians are given by, $H^{t(b)}_\eta = -\hbar v_F \left(\eta q^{t(b)}_{\eta x} \sigma_x+q^{t(b)}_{\eta y}\sigma_y \right)$, where $\mathbf{q}^{t(b)}_{\eta}= \mathcal{R}(\pm \tfrac{\theta}{2})(\mathbf{k}-\mathbf{K}^{t(b)}_\eta)$, with $v_F$ the monolayer graphene Fermi velocity. The interlayer tunneling matrix is :
\begin{equation}
T =
\begin{bmatrix}
u & u' \\
u' & u
\end{bmatrix}
+\begin{bmatrix}
u & u' e^{-i\eta \varphi} \\
u' e^{i\eta \varphi}& u
\end{bmatrix}e^{i \eta \mathbf{G}_1^M \cdot \mathbf{r}}+\begin{bmatrix}
u & u' e^{i\eta \varphi} \\
u' e^{-i\eta \varphi}& u
\end{bmatrix}e^{i \eta (\mathbf{G}_1^M+\mathbf{G}_2^M) \cdot \mathbf{r}},
\end{equation}
 where \(G_j^M\) are defined in Eq.\eqref{eqSM1}, and $\varphi = 2\pi/3$. The $e^{\pm i 2\pi/3}$ factors arise from the threefold ($C_3$) symmetry of graphene and encode the phase differences between the tunneling paths, giving rise to trigonal warping in the band structure. We include lattice distortion effects by taking $u \neq u'$ (with $u = 0.0797\,\text{eV}$ and $u' = 0.0975\,\text{eV}$), which opens energy gaps between the low and higher bands, consistent with experiments. In the limit of vanishing twist angle, which gives \(L_M\rightarrow\infty\), the spatial dependence of $T(\mathbf{r})$ becomes,
\begin{equation}
T =\begin{bmatrix}
3u & u'+ u' e^{-i\eta \varphi}+ u' e^{i\eta \varphi} \\
u'+ u' e^{i\eta \varphi}+ u' e^{-i\eta \varphi} & 3u
\end{bmatrix}=\begin{bmatrix}
3u & u'(1+  2 \cos{(\eta \varphi)})  \\
u'(1+  2 \cos{(\eta \varphi)}) & 3u
\end{bmatrix}=3u\begin{bmatrix}
1 & 0 \\
0 & 1
\end{bmatrix}
\end{equation}
which corresponds to the AA-stacked bilayer graphene limit, where moiré-induced trigonal warping effects are absent \cite{IM6massatt2025defect,IM7hsu2010anomalous}.

The moiré band structure is obtained numerically by diagonalizing the continuum Hamiltonian in a plane-wave basis following Ref.~\cite{koshino2018maximally,37xie2023varphi}. Fig.~\ref{matbgband}(a) shows the band structure of MATBG along the high-symmetry points of the moiré Brillouin zone, highlighting the emergence of flat bands near charge neutrality. Panels (b) and (c) display the lower and upper flat bands, respectively, exhibiting trigonal warping. Fig.~\ref{trigwarp}(a) presents the trigonal-warped Fermi surface of MATBG within the moiré Brillouin zone at $\mu = 0.415~\text{meV}$ (near the high-symmetry point $M_\theta$), lying within the flat-band regime. The color map represents the magnitude of the Fermi velocity. Figure~\ref{trigwarp}(b) shows the same data in polar coordinates, highlighting the anisotropy induced by trigonal warping. The Fermi velocity is computed as $v_f = \sqrt{v_{f_x}^2 + v_{f_y}^2}$, where $v_{f_{x(y)}}$ is the partial derivative of the energy obtained from Eq.\eqref{eqSM2} numerically, at chemical potential \(\mu=0.416 \ \text{meV}\) with respect to \(k_{x(y)}\). Fig.~\ref{aablgband}(a) shows the band structure of AABLG along the high-symmetry points of the monolayer Brillouin zone, while panels (b) and (c) display the corresponding upper bands, which do not exhibit trigonal warping.

\subsection{Blonder-Tinkham-Klapwijk formalism of 1D MATBG-JJs}
\label{SM.2}

To model the junction, we consider an effective one-dimensional Hamiltonian for the $k_y=0$ channel,
$\mathcal{H}_{1D} = \frac{1}{2} \sum_{\eta,\gamma} \int dx \, \Psi_{\eta,\gamma}^\dagger(x)\, \hat{H}_{\eta,\gamma}(x)\, \Psi_{\eta,\gamma}(x)$, where $\eta=\pm$ denotes the valley index and $\gamma=\pm$ labels right- and left-moving states \cite{37xie2023varphi,38hu2023josephson}. The Nambu spinor is $\Psi_{\eta,\gamma}(x) = \left(\psi_{\eta,\gamma}(x), \psi_{-\eta,-\gamma}^\dagger(x)\right)^T$, and the junction Hamiltonian is then,

\begin{equation}
\hat{H}_{\eta,\gamma}(x) =
\begin{bmatrix}
-i\hbar v_{f,\eta,\gamma}(x)\partial_x + \eta \Delta_v(x) & \Delta_s(x) \\
\Delta_s^*(x) & -i\hbar v_{f,-\eta,-\gamma}(x)\partial_x + \eta \Delta_v(x)
\end{bmatrix}.
\end{equation}

The position-dependent Fermi velocity is $v_{f,\eta,\gamma}(x) = v_{s,\eta,\gamma} [\Theta(-x) + \Theta(x-d)] + v_{v,\eta,\gamma}\Theta(x)\Theta(d-x)$, where $v_{s,\eta,\gamma}$ and $v_{v,\eta,\gamma}$ correspond to the superconducting and weak-link regions, respectively. The pairing potential is $\Delta_s(x) = \Delta_s \left(e^{i\phi/2}\Theta(-x) + e^{-i\phi/2}\Theta(x-d)\right)$, with $\phi=\phi_L-\phi_R$, while the valley polarization is $\Delta_v(x) = \Delta_v \Theta(x)\Theta(d-x)$. The scattering states at the left superconductor \(\text{SC}_L\), right superconductor \(\text{SC}_R\) and the MATBG valley-polarized region is given as \cite{37xie2023varphi,38hu2023josephson}:

\begin{equation}
\begin{aligned}
& \psi^L_{s,\eta,\gamma}(x) =
\begin{bmatrix}
e^{-i\gamma \beta} \\ e^{-i\phi/2}
\end{bmatrix}
e^{ik^0_{s,\eta,\gamma}x+\kappa_{\eta,\gamma}x}, \quad x \le 0, \\[6pt]
&\psi_{v,e,\eta,\gamma}(x) =
\frac{1}{\sqrt{N_{e,\eta,\gamma}}}
\begin{bmatrix}
1 \\ 0
\end{bmatrix}
e^{ik_{e,\eta,\gamma}x}, \quad 0 < x < d, \\[6pt]
&\psi_{v,h,\eta,\gamma}(x) =
\frac{1}{\sqrt{N_{h,\eta,\gamma}}}
\begin{bmatrix}
0 \\ 1
\end{bmatrix}
e^{ik_{h,\eta,\gamma}x}, \quad 0 < x < d, \\[6pt]
&\psi^R_{s,\eta,\gamma}(x) =
\begin{bmatrix}
e^{i\gamma \beta} \\ e^{i\phi/2}
\end{bmatrix}
e^{ik^0_{s,\eta,\gamma}(x-d)-\kappa_{\eta,\gamma}(x-d)}, \quad x \ge d.
\end{aligned}
\label{eqSM6}
\end{equation}

Here, $\psi^L_{s,\eta,\gamma}(x)$ and $\psi^R_{s,\eta,\gamma}(x)$ denote the wavefunctions in the left (\(\text{SC}_L\)) and right superconducting (\(\text{SC}_R\)) regions, respectively, while $\psi_{v,e,\eta,\gamma}(x)$ and $\psi_{v,h,\eta,\gamma}(x)$ correspond to the electron- and hole-like states in the weak-link region. The decay constant in the superconducting regions is given by $\kappa_{\eta,\gamma} = \sqrt{\Delta_s^2 - \epsilon_\eta^2}/(\gamma \hbar v_{s,\eta,\gamma})$, with $k^0_{s,\eta,\gamma}$ and $v_{s,\eta,\gamma}$ representing the Fermi wave vector and velocity in the superconducting leads. Phase factor $\beta = \cos^{-1}(\epsilon_\eta/\Delta_s)$ for $|\epsilon_\eta| < \Delta_s$, and $\beta = -i\,\cosh^{-1}(\epsilon_\eta/\Delta_s)$ otherwise \cite{37xie2023varphi,38hu2023josephson}. Time-reversal symmetry in the superconducting leads imposes $k^0_{s,\eta,\gamma} = k^0_{s,-\eta,-\gamma}$. In the weak-link region, $k_{e,\eta,\gamma}$ and $k_{h,\eta,\gamma}$ denote the electron- and hole-like wave vectors. The normalization factors $N_{e(h),\eta,\gamma}$ ensure the scattering matrix to be unitary. To leading order \cite{37xie2023varphi,38hu2023josephson}, $k_{e(h),\eta,\gamma} \approx \gamma k^0_{v,\eta,\gamma} + \delta k_{e(h),\eta,\gamma}$, where $\delta k_{e,\eta,\gamma} = (\epsilon_\eta - \eta \Delta_v)/(\gamma \hbar v_{v,\eta,\gamma})$ and $\delta k_{h,\eta,\gamma} = -(\epsilon_\eta - \eta \Delta_v)/(\gamma \hbar v_{v,-\eta,-\gamma})$. Since the BdG Hamiltonian is block-diagonal in $\eta$, the scattering problem can be solved independently for each valley and the contributions summed to obtain the total Josephson current.

\begin{equation}
\Psi_\eta(x) =
\begin{cases}
a\,\psi^L_{s,\eta,+}(x) + b\,\psi^L_{s,\eta,-}(x), & x \le 0, \\[6pt]
c_e^{+}\,\psi_{v,e,\eta,+}(x) + c_e^{-}\,\psi_{v,e,\eta,-}(x)
+ c_h^{+}\,\psi_{v,h,\eta,+}(x) + c_h^{-}\,\psi_{v,h,\eta,-}(x), & 0 < x < d, \\[6pt]
a'\,\psi^R_{s,\eta,+}(x) + b'\,\psi^R_{s,\eta,-}(x), & x \ge d.
\end{cases}
\end{equation}

To get the ABS energies we apply the boundary condition related to continuity of wavefunction and particle current conservation \cite{37xie2023varphi,38hu2023josephson}, given below:

\begin{equation}
\begin{aligned}
 a\,\psi^L_{s,\eta,+}(0) + b\,\psi^L_{s,\eta,-}(0) &= 
c_e^{+}\,\psi_{v,e,\eta,+}(0) + c_e^{-}\,\psi_{v,e,\eta,-}(0)
+ c_h^{+}\,\psi_{v,h,\eta,+}(0) + c_h^{-}\,\psi_{v,h,\eta,-}(0), \\[6pt]
a'\,\psi^R_{s,\eta,+}(d) + b'\,\psi^R_{s,\eta,-}(d) &= 
c_e^{+}\,\psi_{v,e,\eta,+}(d) + c_e^{-}\,\psi_{v,e,\eta,-}(d)
+ c_h^{+}\,\psi_{v,h,\eta,+}(d) + c_h^{-}\,\psi_{v,h,\eta,-}(d), \\[6pt]
a\,v_{s,\eta,+}\,\psi^L_{s,\eta,+}(0) + b\,v_{s,\eta,-}\,\psi^L_{s,\eta,-}(0) &= 
v_{v,\eta,+}\,c_e^{+}\,\psi_{v,e,\eta,+}(0) + v_{v,\eta,-}\,c_e^{-}\,\psi_{v,e,\eta,-}(0)
- v_{v,-\eta,-}\,c_h^{+}\,\psi_{v,h,\eta,+}(0)
- v_{v,-\eta,+}\,c_h^{-}\,\psi_{v,h,\eta,-}(0), \\[6pt]
a'\,v_{s,\eta,+}\,\psi^R_{s,\eta,+}(d) + b'\,v_{s,\eta,-}\,\psi^R_{s,\eta,-}(d) &= 
v_{v,\eta,+}\,c_e^{+}\,\psi_{v,e,\eta,+}(d)
+ v_{v,\eta,-}\,c_e^{-}\,\psi_{v,e,\eta,-}(d)
- v_{v,-\eta,-}\,c_h^{+}\,\psi_{v,h,\eta,+}(d)
- v_{v,-\eta,+}\,c_h^{-}\,\psi_{v,h,\eta,-}(d).
\end{aligned}
\label{eqSM8}
\end{equation}

For convenience, we define the amplitudes in the left and right superconducting regions as
$a{(L)}=a$, $b{(L)}=b$, $a{(R)}=a'$, $b{(R)}=b'$. The scattering amplitudes are written as
$c_e^{\pm}{ (L)}=c_e^{\pm}$, $c_h^{\pm}{ (L)}=c_h^{\pm}$, while in the right region they acquire phase factors due to propagation,
$c_e^{\pm}{ (R)} = c_e^{\pm} e^{i k_{e,\eta,\pm} d}$ and $c_h^{\pm} {(R)} = c_h^{\pm} e^{i k_{h,\eta,\pm} d}$. We now construct the scattering matrices describing Andreev reflection at the interfaces and propagation through the weak-link region. For simplicity, we employ the Andreev approximation and neglect any chemical potential mismatch at the SC–VP interface, so that $v_{v,\eta,\pm} = v_{s,\eta,\pm}$. The normalization factors in the VP region are taken as $N_{e,\eta,\gamma} = N_{h,\eta,\gamma} = 1$, and only a single clean transport channel is retained in the normal region \cite{37xie2023varphi,38hu2023josephson}
. Solving the linear equations by using Eqs.~\ref{eqSM6} and \ref{eqSM8}, we obtain the following,
\begin{equation}
\psi_{\text{out}} =
\begin{pmatrix}
c_e^{+}(L) \\
c_h^{-}(L) \\
c_e^{-}(R) \\
c_h^{+}(R)
\end{pmatrix}
=
S_A
\begin{pmatrix}
c_e^{-}(L) \\
c_h^{+}(L) \\
c_e^{+}(R) \\
c_h^{-}(R)
\end{pmatrix}
\equiv S_A \psi_{\text{in}},
\qquad \text{where},\quad
S_A =
\begin{pmatrix}
0 & e^{i\phi/2 - i\beta} & 0 & 0 \\
e^{-i\phi/2 - i\beta} & 0 & 0 & 0 \\
0 & 0 & 0 & e^{-i\phi/2 - i\beta} \\
0 & 0 & e^{i\phi/2 - i\beta} & 0
\end{pmatrix}.
\end{equation}

Similarly, the propagation through the normal (weak-link) region is described by,
\begin{equation}
\psi_{\text{in}} =
\begin{pmatrix}
c_e^{-}(L) \\
c_h^{+}(L) \\
c_e^{+}(R) \\
c_h^{-}(R)
\end{pmatrix}
=
S_N
\begin{pmatrix}
c_e^{+}(L) \\
c_h^{-}(L) \\
c_e^{-}(R) \\
c_h^{+}(R)
\end{pmatrix}
\equiv S_N \psi_{\text{out}},
\qquad \text{where},\quad
S_N =
\begin{pmatrix}
0 & 0 & e^{-i k_{e,\eta,-} d} & 0 \\
0 & 0 & 0 & e^{-i k_{h,\eta,+} d} \\
e^{i k_{e,\eta,+} d} & 0 & 0 & 0 \\
0 & e^{i k_{h,\eta,-} d} & 0 & 0
\end{pmatrix}.
\end{equation}

The Andreev bound states (ABS) are obtained from the condition :
\begin{equation}
\det \left( I - S_A S_N \right) = 0,
\end{equation}
which yields,

\begin{equation}
\left(1 - e^{i\phi - 2i\beta - i(\delta k_{h,\eta,+} - \delta k_{e,\eta,+})d}\right)
\left(1 - e^{-i\phi - 2i\beta - i(\delta k_{h,\eta,-} - \delta k_{e,\eta,-})d}\right) = 0.
\label{eqSM14}
\end{equation}

Defining the relevant energy scales as $E_T = \hbar v_T/d$ and $E_A = \hbar v_A/d$ \cite{37xie2023varphi,38hu2023josephson}, where $v_T = 4/(v_{v,+,+}^{-1} + v_{v,+,-}^{-1}+v_{v,-,+}^{-1} + v_{v,-,-}^{-1})$ corresponds to the Thouless velocity scale and $v_A = 2/(v_{v,+,+}^{-1} - v_{v,+,-}^{-1} + v_{v,-,-}^{-1} - v_{v,-,+}^{-1})$ captures the energy scale associated with intra-valley inversion symmetry breaking. The above expression in Eq.\eqref{eqSM14} can be simplified to get the self consistent equation below,

\begin{equation}
\cos\left[2 \left(\beta - \frac{\epsilon_\eta - \eta \Delta_v}{E_T}\right)\right]
=
\cos\left(\phi + \frac{\epsilon_\eta - \eta \Delta_v}{\eta E_A}\right),
\label{eq2:ABS_cond}
\end{equation}

For AABLG, trigonal warping effects are absent and consequently no Fermi-velocity anisotropy is present. As a result, \(v_{v,\eta,-}=v_{v,\eta,+}\), which implies \(E_A \rightarrow \infty\). Therefore, Eq.~\eqref{eq2:ABS_cond} simplifies to

\begin{equation}
    \cos\left[2 \left(\beta - \frac{\epsilon_\eta - \eta \Delta_v}{E_T}\right)\right]
=
\cos\left(\phi \right),
\label{AABLG_eq} 
\end{equation}

To study the thermodynamic cycles in MATBG-JDTM, we set the superconducting phase 
\(\phi=0\) and treat the valley-polarization parameter \(\Delta_v\) as the driving parameter. 
Conversely, for MATBG-JJTM, we set \(\Delta_v=0\) and take the superconducting phase 
\(\phi\) as the driving parameter.

\subsection{Quantum Thermodynamic cycles in Josephson junctions with either superconducting phase (\(\phi\)) or valley polarization potential (\(\Delta_v\)) as system parameters}

\label{SM.3}

\begin{figure}[!htbp]
    \centering
    \includegraphics[width=0.5\linewidth]{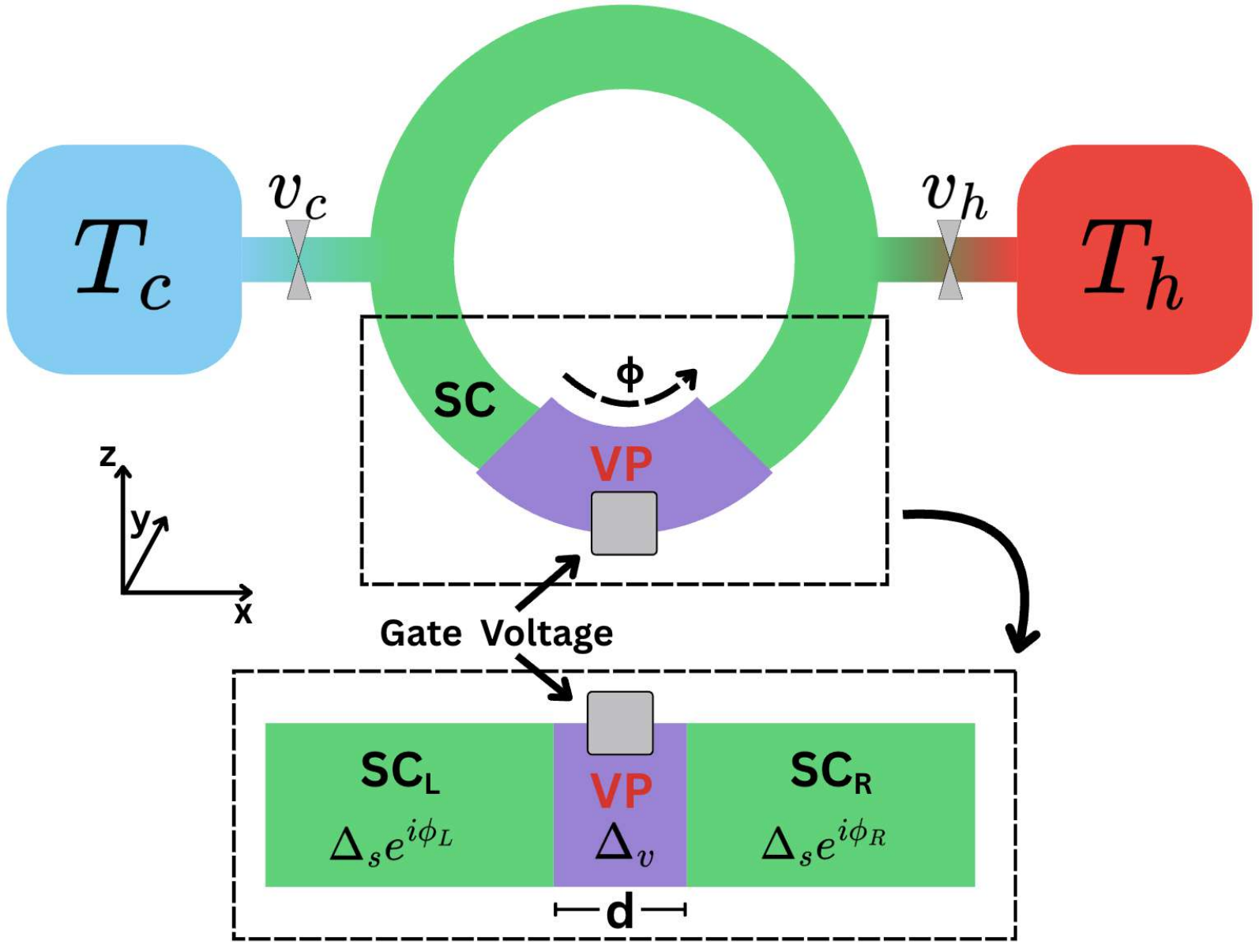}
    \caption{Schematic of a MATBG-based Josephson junction loop with a weak-link valley-polarized region of length $d$, coupled to thermal reservoirs at temperatures $T_h$ and $T_c$ via controllable valves $v_h$ and $v_c$. The junction is characterized by the superconducting order parameter $\Delta_s$, valley-polarization potential $\Delta_v$, and superconducting phase difference $\phi = \phi_L - \phi_R$.}
    \label{figSM5}
\end{figure}

\begin{figure}
    \centering
    \includegraphics[width=1\linewidth]{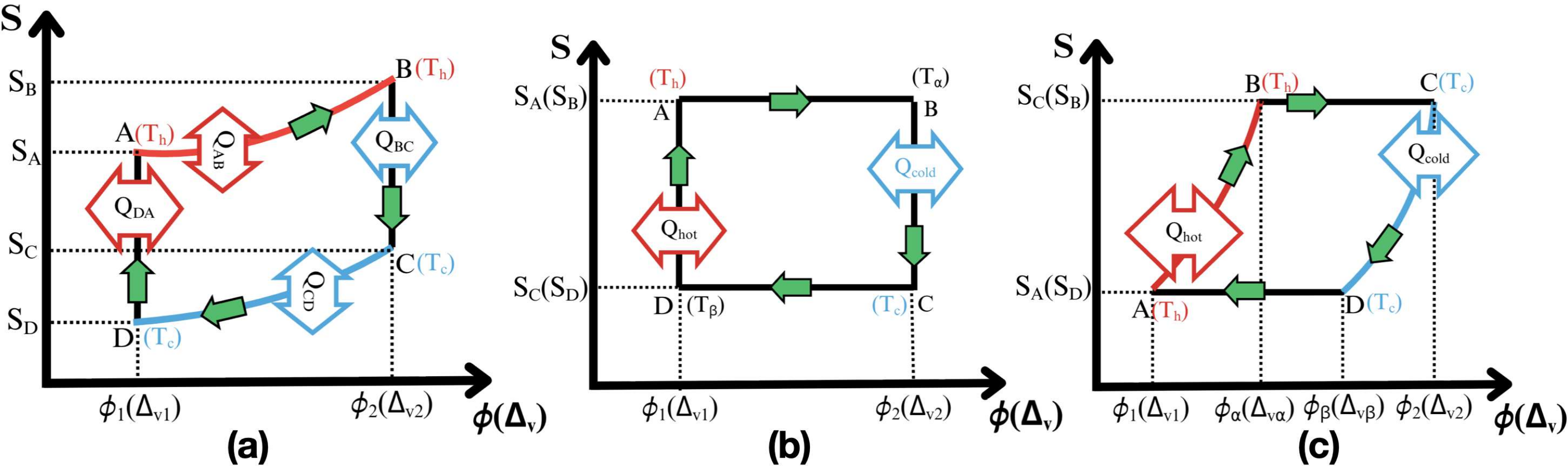}
    \caption{Entropy as a function of system parameters ($\phi,\Delta_v$) for (a) QSC, consisting of two quantum isothermal and two quantum isochoric strokes, (b) QOC, consisting of two quantum adiabatic and two quantum isochoric strokes, and (c) QCC, consisting of two quantum isothermal and two quantum adiabatic strokes. Red arrows indicate heat exchange with the hot reservoir, while blue arrows indicate heat exchange with the cold reservoir.}
    \label{figSM6}
\end{figure}

We consider a general Josephson junction based thermal machine in which either the superconducting phase difference $\phi$ or the valley-polarization potential $\Delta_v$ can act as an external system parameter (see, Fig. \ref{figSM5}). For MATBG-JDTM, we set \(\phi=0\), such that the junction is controlled solely through the valley-polarizing potentials. For MATBG-JJTM and AABLG-JJTM, we set \(\Delta_{v}=0\), allowing the thermodynamic response to be driven entirely by the superconducting phase differences.

Since the change in internal energy over one complete cycle is zero, the first law of thermodynamics requires that the net work and heat exchange balance over the cycle. Adopting the convention that work done by the system is positive and work done on the system is negative, while heat absorbed by the system is positive and heat released is negative, the total work is given by \(W_T = Q_{\text{hot}} + Q_{\text{cold}}\). The corresponding operational regimes and their associated performance metrics for all cycles considered are then determined using Table~I of the main text. Within this framework, we analyze quantum thermodynamic cycles implemented by appropriate modulation of the system parameters and controlled coupling to thermal reservoirs.

\textbf{\textit{Quantum Stirling Cycle (QSC):}} The QSC comprises of two isothermal 
and two isochoric strokes, as illustrated in Fig. \ref{figSM6}(a). Starting from 
$\phi_1(\Delta_{v1})$, the system undergoes an isothermal expansion at $T_h$ with 
valve $v_h$ open, reaching $\phi_2(\Delta_{v2})$ and absorbing heat 
$Q_{AB} = T_h(S_B - S_A)$ from the hot reservoir. Closing $v_h$ and opening $v_c$, 
the system undergoes an isochoric stroke at fixed $\phi_2(\Delta_{v2})$, exchanging 
heat $Q_{BC} = U_C - U_B$ with the cold reservoir. An isothermal compression at $T_c$ 
then returns the system parameter to $\phi_1(\Delta_{v1})$, releasing heat 
$Q_{CD} = T_c(S_D - S_C)$. A final isochoric stroke with $v_c$ closed and $v_h$ 
reopened completes the cycle, restoring the initial thermal state while exchanging heat
$Q_{DA} = U_A - U_D$ with the hot bath. The total heat exchanged with the hot and cold reservoirs is 
$Q_{\text{hot}} = Q_{AB} + Q_{DA}$ and $Q_{\text{cold}} = Q_{BC} + Q_{CD}$, 
respectively.

\textbf{\textit{Quantum Otto Cycle (QOC):}} The QOC consists of two adiabatic and 
two isochoric strokes, as shown in Fig. \ref{figSM6}(b). Starting from 
$\phi_1(\Delta_{v1})$ at $T_h$ with both reservoirs decoupled, the system undertakes
a slow adiabatic stroke and reaches $\phi_2(\Delta_{v2})$, which is at an intermediate 
temperature $T_\alpha$ satisfying the isentropic constraint 
$S_A(\phi_1(\Delta_{v1}),T_h) = S_B(\phi_2(\Delta_{v2}),T_\alpha)$. An isochoric 
stroke with $v_c$ open then thermalizes the system to $T_c$, while exchanging heat 
$Q_{\text{cold}} = U_C - U_B$ with the cold bath. A second adiabatic stroke restores the system parameter to $\phi_1(\Delta_{v1})$, along with attaining a relaxation temperature $T_\beta$ via satisfying the adiabatic constrain :
$S_C(\phi_2(\Delta_{v2}),T_c) = S_D(\phi_1(\Delta_{v1}),T_\beta)$. A final isochoric 
stroke with $v_h$ open returns the system to $T_h$, exchanging heat 
$Q_{\text{hot}} = U_A - U_D$ with the hot bath. Unlike the QSC, the QOC involves four 
distinct temperatures $T_h$, $T_c$, $T_\alpha$, and $T_\beta$, all of which must 
remain below the superconducting critical temperature for the cycle to be valid.

\textbf{\textit{Quantum Carnot Cycle (QCC):}} The QCC consists of two isothermal 
and two adiabatic strokes, as illustrated in Fig. \ref{figSM6}(c). Starting from 
$\phi_1(\Delta_{v1})$ at $T_h$, an isothermal stroke with $v_h$ open drives the 
system to $\phi_\alpha(\Delta_{v\alpha})$, exchanging heat 
$Q_{\text{hot}} = T_h(S_B - S_A)$ with the hot bath. A subsequent adiabatic stroke with both valves 
closed brings the system to $\phi_2(\Delta_{v2})$ at $T_c$, satisfying 
$S_B(\phi_\alpha(\Delta_{v\alpha}),T_h) = S_C(\phi_2(\Delta_{v2}),T_c)$. An 
isothermal compression at $T_c$ with $v_c$ open then drives the system to 
$\phi_\beta(\Delta_{v\beta})$, releasing heat $Q_{\text{cold}} = T_c(S_D - S_C)$ with the cold bath. 
A final adiabatic stroke with both valves closed restores the initial state 
$\phi_1(\Delta_{v1})$ at $T_h$, satisfying the condition 
$S_D(\phi_\beta(\Delta_{v\beta}),T_c) = S_A(\phi_1(\Delta_{v1}),T_h)$.

\subsection{Fermi--Dirac Distribution and Entropy for MATBG-JDTM, MATBG-JJTM, and AABLG-JJTM}
\label{SM.4}

In this section we calculate the Fermi--Dirac occupation probabilities $f_{\eta,n}$ for the three junctions. The occupation of the $n$-th ABS in valley $\eta$ is given by \cite{24vischi2019thermodynamics}:
\begin{equation}
    f_{\eta,n}(\Delta_v,\phi,T) = \frac{1}{e^{\,\epsilon_{\eta,n}(\Delta_v,\phi)/k_B T}+1}.
    \label{fermi_EQ}
\end{equation}
\begin{figure}[!htbp]
    \centering
    \includegraphics[width=\linewidth]{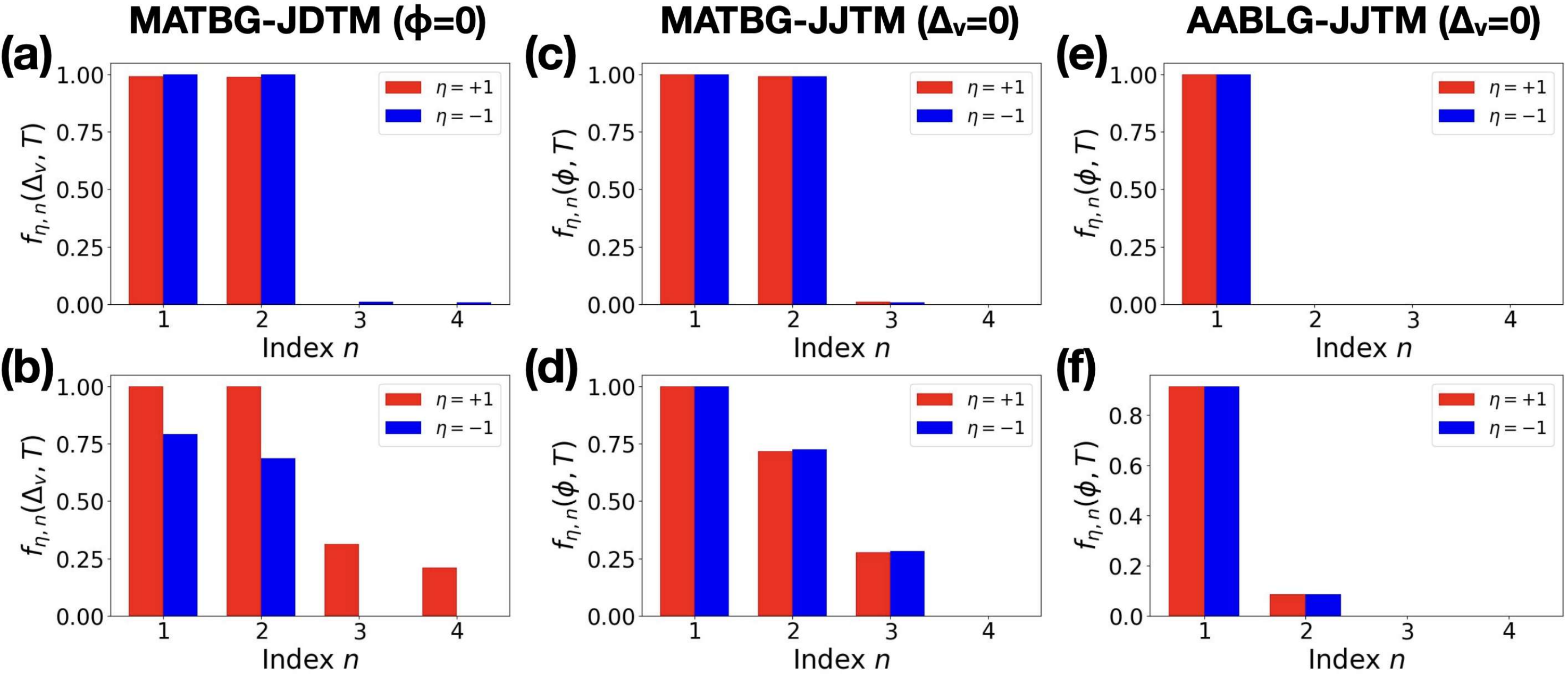}
    \caption{Fermi--Dirac occupation probabilities, \(f_{\eta,n}\), displayed as bar charts for the individual ABS levels indexed by \(n\). Panels (a,b) correspond to MATBG-JD at \(\phi=0\), for (a) \(\Delta_v = 0.052~\text{meV}\) and (b)\(0.113~\text{meV}\). Panels (c,d) show MATBG-JJ with \(\Delta_v=0\) at phase differences (c)\(\phi=\pi/2\) and (d)\(\phi=1.1\pi\). Panels (e,f) show AABLG-JJ with \(\Delta_v=0\) at phase differences (e)\(\phi=\pi/2\) and (f)\(\phi=1.1\pi\). In all cases, the temperature is fixed at \(T=0.075~\text{K}\).}
    \label{fig:fermi_bar}
\end{figure}

With the parameters $E_T = 0.065~\text{meV}$ and $E_A = 1.2~\text{meV}$ chosen to reproduce the ABS spectrum of MATBG (see Eq.~\eqref{eq2:ABS_cond}), each valley index supports up to four ABS energies for both MATBG-JJ and MATBG-JD ($n_\text{max} = 4$) 
\cite{37xie2023varphi}. AABLG has only two ABS per valley ($E_T = 5.75~\text{meV}$, $E_A \to \infty$). Consequently, as seen in Fig.~\ref{fig:fermi_bar}(a)--(d), MATBG-JD and MATBG-JJ exhibit finite occupation probabilities up to higher level indices across a wider region of parameter space, compared to AABLG-JJ in Fig.~\ref{fig:fermi_bar}(e)--(f). This larger number of thermally active levels gives rise to finite entropy over a broader parameter range, which in turn enables finite work output and heat exchange over a wider domain of the thermodynamic cycle.

Entropy is calculated from the standard expression \cite{IM2quan2007quantum,24vischi2019thermodynamics}:
\begin{equation}
    S(\Delta_v,\phi,T) = -k_B \sum_{\eta,n}
    \Bigl[f_{\eta,n} \ln f_{\eta,n} + (1-f_{\eta,n})\ln(1-f_{\eta,n})\Bigr].
    \label{ENTROPY_EQ}
\end{equation}
\begin{figure}[!htbp]
    \centering
    \includegraphics[width=\linewidth]{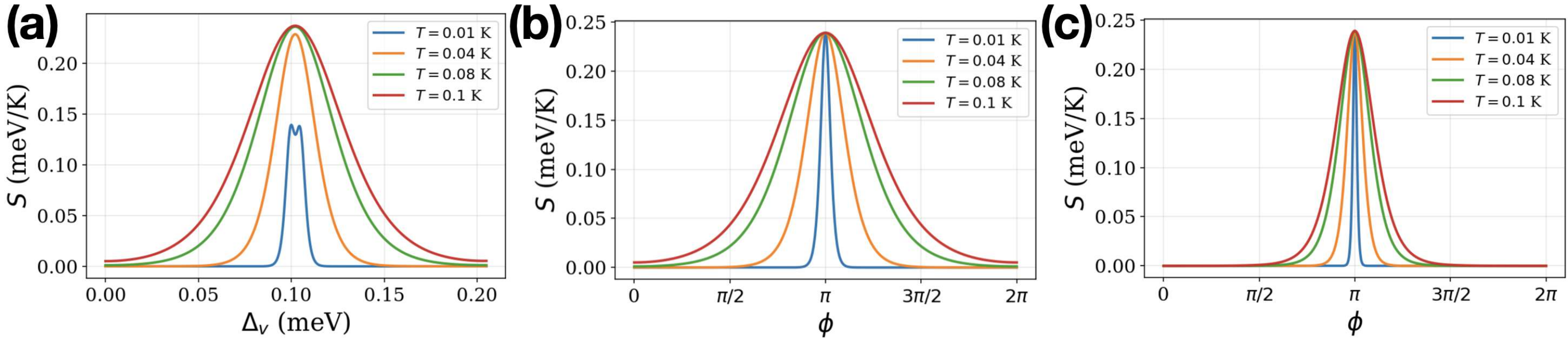}
    \caption{Entropy $S$ at different temperatures for
    (a)~MATBG-JDTM with $\Delta_v$ as the tuning parameter at $\phi=0$,
    (b)~MATBG-JJTM with $\phi$ as the tuning parameter at $\Delta_v=0$, and
    (c)~AABLG-JJTM with $\phi$ as the tuning parameter at $\Delta_v=0$. }
    \label{fig:entropy_params}
\end{figure}

Fig.~\ref{fig:entropy_params} shows the entropy for each junction. The MATBG-based devices exhibit a broad entropy distribution across the tuning-parameter space, whereas in AABLG-JJTM the entropy is appreciable only in the vicinity of $\phi \approx \pi$. Since the heat exchanged in a thermodynamic cycle is directly related to the change in entropy, this broader entropy landscape translates into finite heat exchange over a much wider operating range in MATBG-based devices, making them more favourable platforms for realizing thermal machines. 
In MATBG-JDTM, the thermodynamic cycle is driven by $\Delta_v$ via an electrostatic gate, avoiding the flux noise associated with cycling $\phi$ through a complete cycle in MATBG-JJTM, highlighting a potential experimental advantage of the MATBG-JDTM over MATBG-JJTM.

\subsection{Adiabatic constraints and relaxation temperatures in the QOC}
\label{SM.5}

\begin{figure}[!htbp]
    \centering
    \includegraphics[width=0.9\linewidth]{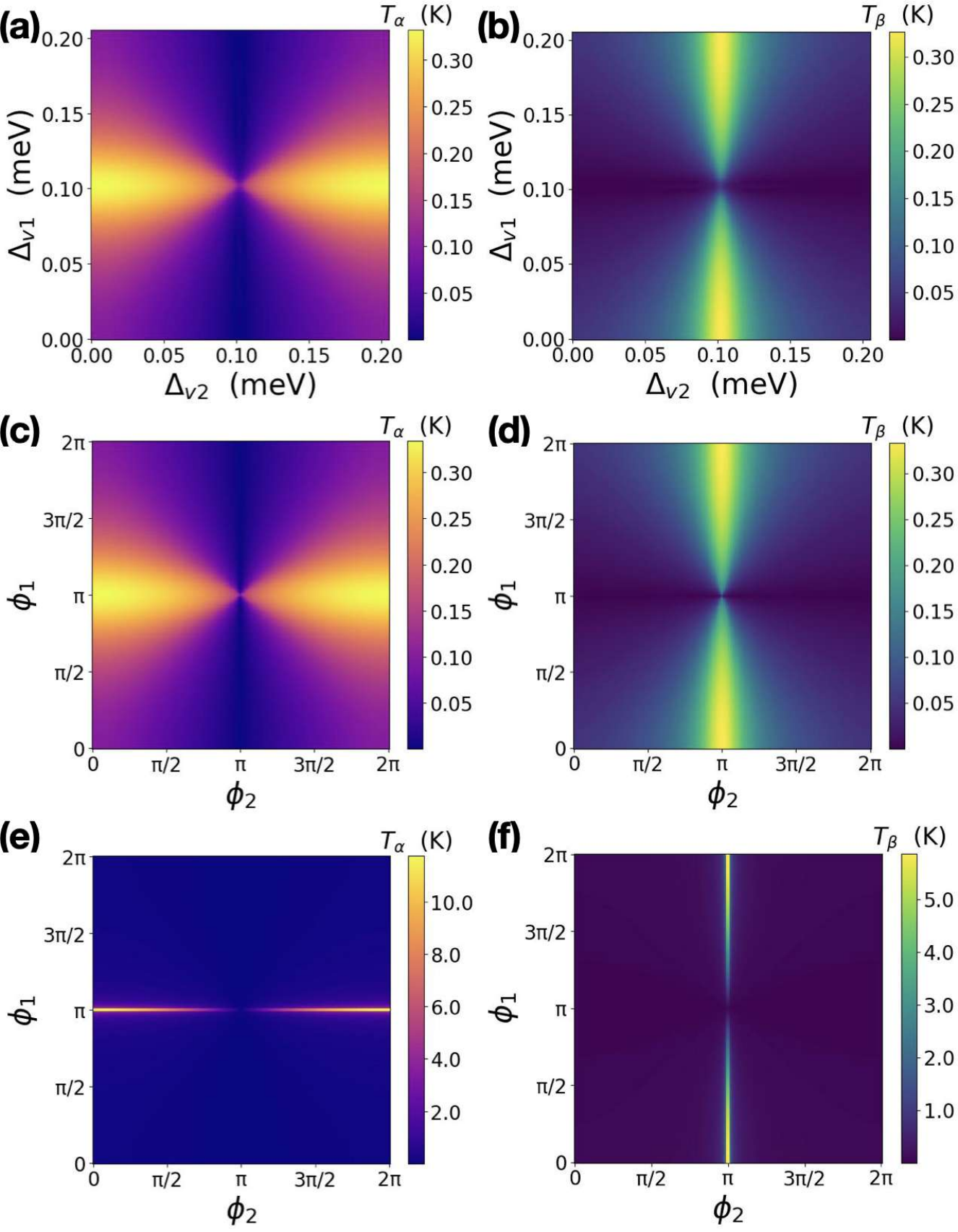}
    \caption{(a,b) Contour plots of relaxation temperatures $T_\alpha$ and 
    $T_\beta$ for MATBG-JDTM, (c,d) for MATBG-JJTM, and (e,f) for 
    AABLG-JJTM, across the respective control parameter spaces. Regions where 
    the relaxation temperatures diverge or exceed the superconducting critical 
    temperature are thermodynamically inaccessible and must be excluded from 
    the operational regime of the Otto cycle.}
    \label{fig:relax}
\end{figure}

In the QOC, the two adiabatic strokes impose isentropic 
constraints that uniquely determine a pair of intermediate relaxation 
temperatures, $T_\alpha$ and $T_\beta$, at the end of each adiabatic process. 
These temperatures are not externally set but instead emerge self-consistently 
from the junction's ABS spectrum and the thermodynamic state of the system 
at the beginning of each adiabatic stroke. For JDTM, where $\Delta_v$ serves 
as the external control parameter at fixed $\phi$, the constraints take the form:
\begin{equation}
S(\Delta_{v1}, T_h) = S(\Delta_{v2}, T_\alpha), \qquad
S(\Delta_{v2}, T_c) = S(\Delta_{v1}, T_\beta),
\label{entropy_cond1}
\end{equation}
while for JJTM the analogous constraints are written with $\phi$ as the control parameter:

\begin{equation}
S(\phi_1, T_h) = S(\phi_2, T_\alpha), \qquad
S(\phi_2, T_c) = S(\phi_1, T_\beta),
\label{entropy_cond2}
\end{equation}

Here, $T_\alpha$ and $T_\beta$ are the relaxation temperature attained after the adiabatic processes. For the QOC to be thermodynamically consistent, all four temperatures $T_h$, $T_c$, $T_\alpha$, and $T_\beta$ must remain below the superconducting critical temperature $\theta_c$, since the entire Josephson junction description rests on the existence of a well-defined superconducting condensate. To understand how these relaxation temperatures are distributed across the parameter space, we solve the isentropic constraints in Eqs.\eqref{entropy_cond1} and \eqref{entropy_cond2} numerically at each  point and present the results as contour maps in Fig.~\ref{fig:relax}.

Entropy $S$, whether parametrised by $\Delta_v$ or $\phi$, is directly determined by the ABS spectrum through Eqs.~\eqref{fermi_EQ} and \eqref{ENTROPY_EQ}, and entropy is maximised when either temperature \(T_\alpha\) or \(T_\beta\) tends to infinity or the ABS energies \(\epsilon_{\eta,n}\) vanish, since both conditions drive the Fermi--Dirac occupation toward $f_{\eta,n} \to 1/2$. For MATBG-JDTM and MATBG-JJTM, the ABS energies remain finite and well-separated from zero across the entire operational parameter space (see, Fig.~6 in EM). The isentropic constraints can therefore always be satisfied at physically reasonable temperatures as shown in Fig.~\ref{fig:relax}(a--d), with $T_\alpha$ and $T_\beta$ remaining below the superconducting critical temperature $\theta_c$ throughout, confirming that the QOC is thermodynamically well-posed for these platforms across their full operational regimes. In contrast, for AABLG-JJTM near $\phi \approx \pi$, the ABS energies approach zero, $\epsilon_{\eta,n} \to 0$, so the Fermi--Dirac occupation saturates to $f_{\eta,n} \to 1/2$, corresponding to maximum entropy $S_{\max}$. Satisfying an isentropic constraint of the form $S(\phi_1 \approx \pi,\, T_1) = S(\phi_2,\, T_2)$ then requires $T_2 \to \infty$ wherever the ABS energies are non-zero, causing the relaxation temperature to diverge. This divergence manifests as a streak of yellow-red in Fig.~\ref{fig:relax}(e) along \(\phi_1\) and yellow-green in Fig.~\ref{fig:relax}(f) along \(\phi_2\), signals the breakdown of the superconducting condensate, and renders the corresponding region of parameter space physically inaccessible. This corresponds to the blackened region in Fig. 4(e,f) in the main text at \(\phi_1,\phi_2 \approx \pi\). Consequently, operation of AABLG-JJTM as a QOC must be restricted to regimes sufficiently far from $\phi \approx \pi$, where the ABS energies remain finite and the relaxation temperatures stay within $\theta_c$.

\subsection{Algorithm used to obtain the performance of  thermal machines in MATBG-JDTM, MATBG-JJTM and AABLG-JJTM}
\label{SM.6}

We outline below the step-by-step numerical procedure used to evaluate the thermodynamic performance of MATBG and AABLG junction based thermal machines studied in this work.

\textbf{Step 1: Solving the ABS condition.} We numerically solve the ABS condition in  Eq.~\eqref{eq2:ABS_cond}, for appropriate values of the characteristic energy scales $E_A$ and $E_T$ \cite{37xie2023varphi}, which encode the junction's Thouless energy and the intra-valley inversion-symmetry-breaking scale, respectively. For JDTM, the ABS energies $\epsilon_{\eta,n}$ are obtained as a function of the valley-polarization potential $\Delta_v$ at phase difference $\phi=0$, while for JJTM they are obtained as a function of $\phi$ at $\Delta_v=0$ for each valley 
index $\eta = \pm$, which together constitute the ABS spectrum of the junction.

\textbf{Step 2: Computing thermodynamic quantities.} Using the ABS spectrum obtained in Step 1, we compute the equilibrium thermodynamic quantities like free energy, entropy, and internal energy (see, Eq. 3 in the main text) depending on the relevant parameters. These quantities form the thermodynamic state functions that serve as inputs to the thermodynamic cycle design in Step 3.

\textbf{Step 3: Constructing the thermodynamic cycle.} Given the thermodynamic state functions over the full parameter space, we implement the appropriate quantum thermodynamic cycle, QSC, QOC, or QCC, by executing the corresponding sequence of strokes as 
described in Sec.~\ref{SM.3}. For each stroke, the heat exchanged is evaluated at every point in the parameter space according to the stroke type: isothermal strokes imply $Q = T\,\Delta S$, isochoric strokes imply $Q = \Delta U$ at fixed control parameters, and adiabatic strokes satisfy $\Delta S = 0$. The work output per cycle is then obtained from the first law.

\textbf{Step 4: Identifying operational phases and performance metrics.} 
From the heat and work computed in Step 3, we classify each point in the parameter space according to its operational phase, heat engine, refrigerator, cold pump, or Joule pump, based on the signs of $W_T$, $Q_{\text{hot}}$, and $Q_{\text{cold}}$ (see Table. 1 in main text), and associated performance metrics are then evaluated. This entire procedure is repeated across different material platforms (MATBG and AABLG) and junction parameters to enable a systematic comparative analysis  of thermodynamic performance.

The Python code used to generate plots in the main text, EM and SM are uploaded to GitHub \cite{Code}.

\end{document}